\documentclass[12pt]{article}
\usepackage[symbol]{footmisc}
\def\correspondingauthor{\footnote{Corresponding author.  }}
\usepackage[left=2cm,top=2.5cm,right=2cm,bottom=2.5cm]{geometry}
\usepackage{amsmath, amssymb}

\usepackage{graphicx}
\usepackage{graphics}
\usepackage{epstopdf}
\usepackage{subfigure}
\usepackage{amsfonts}
\usepackage{sectsty}
\usepackage{sectsty}
\usepackage{hyperref}
\usepackage{cite}

\begin{document}
	
	\begin{center}
	\large{\bf{Traversable Wormholes in $R+\alpha R^n$ Gravity}} \\
	\vspace{5mm}
	\normalsize{Nisha Godani$^1$ and Gauranga C. Samanta$^{2}$\correspondingauthor{} }\\
	\normalsize{$^1$Department of Mathematics, Institute of Applied Sciences and Humanities\\ GLA University, Mathura, Uttar Pradesh, India }\\
	$^2$Department of Mathematics, BITS Pilani K K Birla Goa Campus, Goa, India \\
	\normalsize {nishagodani.dei@gmail.com \\gauranga81@gmail.com}
\end{center}
\begin{abstract}
In this work, the study of traversable wormholes in
$f(R)$ gravity with the function $f(R)=R+\alpha R^n$, where $\alpha$ and $n$ are arbitrary constants, is taken into account.
The shape function $b(r)=\frac{r}{\exp(r-r_0)}$, proposed by Samanta et al. \cite{godani1}, is considered.  The energy conditions with respect to both constant and variable redshift functions are discussed and the existence of wormhole solutions without presence of exotic matter is investigated.
\end{abstract}

\section{Literature Survey}
Traversable wormholes can be interpreted as hypothetical structures that allow the observers to traverse freely through the throat. These structures appear as a tool connecting two different space-times or two different locations of the same space-time. The study of wormholes was initiated by Flamm \cite{flamm}. After that, Einstein \& Rosen \cite{eins-ros} developed Einstein-Rosen bridge connecting two asymptotically flat space-times. Morris \& Thorne \cite{morris1} proposed traversable wormholes for the fast interstellar travel of an observer through the space-time.  Wormholes are obtained as classical solutions of Einstein's gravitational field equations which exist in the presence of a matter with negative energy called exotic matter. The exotic matter includes a stress energy tensor that does not satisfy the null energy condition (NEC). Hochberg and Visser \cite{Hochberg2} also proved the violation of NEC as a common feature for a static wormhole in general relativity. Later on, they extended this result for dynamic wormhole \cite{Hochberg1}.  Several cosmologists have tried to explore the stability of wormholes  and find the ways to avoid or minimize the violation of NEC.
Shinkai and Hayward \cite{Shinkai} studied the stability of traversable wormholes.
Bergliaffa and  Hibberd \cite{Bergliaffa} examined the stress-energy tensor and obtained that the rotating wormhole can be explained neither by a perfect fluid nor by a fluid with anisotropic stress.
Kuhfittig \cite{Kuhfittig} using time dependent angular velocity studied  rotating axially symmetric wormholes as a generalization of static and spherically symmetric traversable wormhole. They also analyzed the effect of angular velocity on weak energy condition in case of both axially and spherically symmetric wormholes.
Ayg\"{u}n et al. \cite{aygun} studied rigidly rotating wormhole in the background of Einstein's general theory of relativity.
B\"{o}hmer et al. \cite{Bohmer} used a linear dependence between energy density and pressure and studied  wormhole solutions.
 Bronnikov and Galiakhmetov \cite{Bronnikov1} using the framework of Einstein-Cartan theory studied the  existence  of  static  traversable  wormholes  without   exotic  matter.
 Wang and Meng \cite{wang} obtained  wormhole solutions in the framework of bulk viscosity using three classes of viscous  models.
  Moradpour \cite{moradpour} investigated traversable wormholes in both Einstein's general relativity and Lyra geometry.
  Barros and Lobo \cite{barros} obtained  wormhole solutions using three form fields and analyzed the validation of weak and null energy conditions.
  Tsukamoto and Kokubu \cite{kokubu} studied stability of thin shell wormholes in the presence of barotropic fluid.\\
  
\noindent
 The latest astronomical observations suggest that the expansion of the universe is in accelerating way\cite{Riess, perl}. This acceleration is driven by a gravitationally repulsive energy called dark energy. For the explanation of this phenomenon, several proposals have been proposed during last two decades. One of them is the modification of the Einstein's general relativity by modifying the Lagrangian gravitational action $R$, where $R$ is the Ricci scalar curvature. Several theories are introduced in literature that modifies Einstein's action. A significant theory that explains the cosmic acceleration and other cosmological issues is $f(R)$ theory of gravity. In this theory, the Einstein's gravitational action is replaced by a general function of $R$, $f(R)$. For the particular case $f(R)=R$, the $f(R)$ theory is equivalent to general relativity. The idea behind this generalization is that the results may vary with the variation of choice of function $f(R)$. Earlier the exploration of inflation scenario created interest in $f(R)$ theory and Starobinsky \cite{Star} provided an inflation model using this theory.  Nojiri and Odintsov \cite{Nojiri} defined an $f(R)$ model with positive and negative powers of $R$ supporting early inflation and late time acceleration. Carroll et al. \cite{Carroll} showed the early and  late time accelerations by doing some tiny modifications in action of general relativity.
  Lin et al. \cite{Ting} examined the local and cosmological tests for $f(R)$ theory by several observations. Various other $f(R)$ models are also developed and studied \cite{sotiriou, huang, noj, cog2008, felice, Bamba4, Bamba3, Bamba1, sebas, thakur, Bamba, Peng, Motohashi, Astashenok, noj2017, baha1, Yousaf, Faraoni, Abbas, Sussman, Mongwane, Motohashi12, Mansour, Muller, Wang, Papagiannopoulos, Oikonomou, Chakraborty, Nashed, Capozziello, Gu, Abbas1, Mishra, Odintsov, Castellanos,nginjp}.\\
  
\noindent
The study of wormholes is also extended using $f(R)$ theory.
  Lobo and Oliveira \cite{Lobo1} investigated traversable wormholes using the framework of $f(R)$ gravity and obtained the factors responsible for the violation of null energy condition and supporting the existence of wormholes. They also obtained wormhole solutions for different shape functions.
Bronnikov et al. \cite{Bronnikov2} discussed the existence of wormholes in scalar tensor theory and $f(R)$ gravity.
  Saiedi and Esfahani \cite{Saiedi} studied null and weak energy conditions for wormholes with  constant shape and redshift functions in $f(R)$ gravity.
 Bahamonde et al. \cite{baha} developed dynamical wormholes in $f(R)$ gravity.
 Peter \cite{peter} computed wormhole solutions using different types of shape functions in $f(R)$ gravity and studied traversable wormholes.
Kim\cite{kim} studied FRW model with traversable wormhole. They proved that the violation of energy conditions is not necessary by the total matter in the cosmological model.
Hochberg et al. \cite{Hochberg} solved semi classical field equations representing wormholes.
Lemos et al. \cite{lemos} studied static and spherically symmetric traversable wormholes in the presence of cosmological constant. They explored the properties of traversable wormholes due to the presence of cosmological term.
H. Maeda and  M.  Nozawa \cite{Maeda} studied the properties of $n$-dimensional static wormhole solutions, where $n\geq5$, using Einstein-Gauss-Bonnet gravity.
 Celis et al. \cite{Celis} studied thin shell wormholes using theories beyond general relativity in greater than 4-dimensional space-time.
Rahaman et al. \cite{Rahaman} obtained various wormhole solutions using Finslerian structure of space-time.
Zubair et al. \cite{zubair} assumed fluids of three types and explored energy conditions for static and spherically symmetric wormholes in $f(R,\phi)$ gravity.
Godani and Samanta \cite{godani} studied traversable wormhole in $f(R)$ gravity and explored energy conditions using two shape functions. Samanta et al. \cite{godani1} defined a new shape function and studied energy conditions in both $f(R)$ and $f(R,T)$ theories. Recently, Godani and Samanta \cite{ngmpla} and Samanta and Godani \cite{gnmpla} investigated energy conditions for traversable wormholes in $f(R)$ gravity.\\

\noindent
The aim of this paper is to study the wormhole geometry equipped with minimum amount of exotic matter near the throat in $f(R) = R+\alpha R^n$ gravity with shape function $b(r)=\frac{r}{\exp(r-r_0)}$. This shape function was introduced by Samanta et al. \cite{godani1} to compare the wormhole solutions in $f(R)$ and $f(R,T)$ gravity theories. They found the satisfaction of energy conditions for small range of $r$ in $f(R)$ gravity with $f(R)=R-\mu R_c\tanh\frac{R}{R_c}$, where $\mu$ and $R_c$ are constants, and for a wide range of $r$ in $f(R,T)$ theory of gravity with function $f(R,T)=R+2\lambda T$, where $\lambda$ is an arbitrary constant. However, there may be possibility of getting the validation of energy conditions for wide range of $r$ or there may be possibility of wormholes with some significant geometric configuration, in the framework of $f(R)$ gravity with some different choice of  $f(R)$ function. To explore these possibilities, we have considered the same shape function with different $f(R)$ function defined as $f(R)= R+\alpha R^n$, where $\alpha$ and $n$ are constants. Further, Samanta et al. \cite{godani1} considered constant redshift function to investigate wormhole solutions. Anchordoqui et al. \cite{Anchordoqui} considered variable redshift function $\Phi(r)=-\frac{\alpha}{r}$, $\alpha>0$ and obtained analytical solutions explaining wormhole geometries. Sarkar et al. \cite{Sarkar} assumed $\Phi(r)=\frac{\alpha}{r}$, where $\alpha$ is a constant, to explore wormhole solutions in $\kappa(R,T)$ gravity and obtained wormhole solutions filled with exotic type matter everywhere. Rahaman et al. \cite{Rahaman} used two forms of redshift function (i) $\Phi(r)=\frac{\alpha}{r}$ and (ii) $\Phi(r)=\ln(\frac{\sqrt{\gamma^2++r^2}}{r})$,  where $\alpha$ and $\gamma$ are constant. Using these forms of redshift functions with specific choice of shape function, they determined generating functions comprising  the wormhole like geometry. Further using $\Phi(r)=\frac{\alpha}{r}$ with particular form of generating function, they derived shape function of wormhole solutions. Further, Pavlovic and  Sossich  \cite{Pavlovic} investigated possible wormhole solutions in the context of four viable $f(R)$  models, namely the MJWQ model \cite{Miranda}, the exponential gravity model \cite{Cognola, Elizalde}, the Tsujikawa model \cite{Tsujikawa, Felice} and the Starobinsky model \cite{Tsujikawa, Starobinsky, Amendola1, Amendola2, Amendola3}. They proposed the redshift function $\Phi(r)=\ln(\frac{r_0}{r}+1)$ and  for first three models, they obtained wormhole solutions without need of exotic matter which is a very significant result. It could be possible because of the suitable choice of redshift function.  This work of Pavlovic and  Sossich  \cite{Pavlovic} inspires to check the validity of energy conditions in other $f(R)$ models and hence to examine the type of matter required to sustain the wormhole solutions. Therefore, we have taken the same redshift function $\Phi(r)=\ln(\frac{r_0}{r}+1)$,  in the present work, to analyze the energy conditions and investigate the wormhole solutions.
 The section-wise description is as follows: In section-2, field equations and wormhole geometry is presented. In section-3, solutions of the wormhole geometry are presented. In section-4, various energy conditions are discussed. Results obtained are presented in section-5. Finally, conclusions are provided in section-6.

\section{Field Equations and Wormhole Geometry}
The four dimensional Einstein Hilbert action with matter content of the universe can be written as
\begin{equation}\label{action}
  S=\int\left(\sqrt{-g}\frac{R}{2\kappa^2}+L_m\right)d^4x,
\end{equation}
where $c=1$ and $\kappa^2=8\pi G$, $R$ is the Ricci scalar, $g$ is the determinant of the metric tensor and $L_m$ is the lagrangian for the matter part of the universe.

\noindent
The Einstein field equations can be obtained by varying the action \eqref{action} with respect to the metric tensor $g_{\mu\nu}$.
The Einstein Hilbert action plays an important role for modifying geometry. Hence, the modified gravity can be obtained by modifying the Einstein Hilbert
action. Therefore, to obtain $f(R)$ gravity, we have to replace $f(R)$ in place of $R$ in Einstein Hilbert action. Hence, the four dimensional modified Einstein Hilbert action with matter content for $f(R)$ gravity can be written as
\begin{equation}\label{fr action}
  S=\int\left(\sqrt{-g}\frac{f(R)}{2\kappa^2}+L_m\right)d^4x.
\end{equation}
Now, varying the action \eqref{fr action} with respect to $g_{\mu\nu}$, we can have
\begin{equation}\label{field}
  f^{'}(R)R_{\mu\nu}-\frac{1}{2}f(R)g_{\mu\nu}=\bigtriangledown_{\mu}\bigtriangledown_{\nu}f^{'}(R)-g_{\mu\nu}\bigtriangledown_{\sigma}
  \bigtriangledown^{\sigma}f^{'}(R)+\kappa^2T_{\mu\nu},
\end{equation}
where prime denotes the derivative with respect to the scalar curvature $R$, i.e. $f^{'}=\frac{df}{dR}$. The trace of equation \eqref{field} gives
\begin{equation}\label{}
  3\square f^{'}(R)+Rf^{'}(R)-2f(R)=\kappa^2T,
\end{equation}
where $\square\equiv g^{\mu\nu}\bigtriangledown_{\mu}\bigtriangledown_{\nu}$, $T=g^{\mu\nu}T_{\mu\nu}$ is the trace of the energy momentum tensor $T_{\mu\nu}$ and the Ricci scalar $R=g^{\mu\nu}R_{\mu\nu}$. The field equations \eqref{field} can be written in the following form
\begin{equation}\label{}
  G_{\mu\nu}=\chi_{eff}\left(T_{\mu\nu}+T_{\mu\nu}^{f(R)}\right),
\end{equation}
where $G_{\mu\nu}=R_{\mu\nu}-\frac{1}{2}g_{\mu\nu}$, $\chi_{eff}=\frac{\kappa^2}{f^{'}(R)}$ and $T_{\mu\nu}^{f(R)}$ could be observed as
effective energy momentum tensor in modified $f(R)$ gravity, which is expressed as
\begin{equation}\label{}
  T_{\mu\nu}^{f(R)}=(\bigtriangledown_{\mu}\bigtriangledown_{\nu}-g_{\mu\nu}\bigtriangledown_{\sigma}\bigtriangledown^{\sigma})f^{'}(R)
  +\frac{1}{2}(f(R)-Rf^{'}(R))g_{\mu\nu}.
\end{equation}
For $f(R)=R$, the aforesaid modified theory reduces to general relativity.

\noindent
In this paper, we have considered a spherically symmetric and static wormhole metric \cite{morris1}, which is defined as
\begin{equation}\label{metric}
ds^2=-e^{2\Phi(r)}dt^2+\frac{dr^2}{1-\frac{b(r)}{r}} + r^2(d\theta^2+\sin^2\theta d\phi^2).
\end{equation}
The metric function  $\Phi(r)$ is related to gravitational redshift and  $b(r)$ determines the shape of the wormholes \cite{morris1, Visser, fsnlobo}. Hence $b(r)$ and $\Phi(r)$ are, respectively, called the shape and redshift functions of radial coordinate $r$ that varies from $r_0$ to $\infty$, where $r_0$ is known as radius of the throat. At the throat of the wormhole, the shape function must satisfy $b(r_0)=r_0$. The metric coefficient $g_{rr}$ becomes
infinity at the throat, which is gestured by the coordinate singularity. The proper radial distance
$l(r)=\pm\int_{r_0}^{r}\left(1-\frac{b(r)}{r}\right)^{-\frac{1}{2}}dr$ is required to be finite everywhere. The absence of horizons is necessary for traversable wormhole. This implies that $e^{2\Phi(r)}\ne 0$, so $\Phi(r)$ must be finite everywhere.
Another interesting feature of the redshift function is: the derivative of the redshift function with respect to radial coordinate determines the attractive or repulsive nature of the wormhole geometry.
Since our metric is spherically symmetric, so without loss of generality, one may consider an equatorial slice $\theta=\frac{\pi}{2}$ and for a fixed moment of time i.e. $t=$ constant, the metric \eqref{metric}  becomes
\begin{equation}\label{remetric}
  ds^2=\frac{dr^2}{1-b(r)/r}+r^2d\phi^2.
\end{equation}
The equation \eqref{remetric} can be written in cylindrical coordinates, $(r, \phi, z)$ as
\begin{equation}\label{remetric1}
  ds^2=dz^2+dr^2+r^2d\phi^2.
\end{equation}
In the three-dimensional Euclidean space the embedded surface has equation $z=z(r)$, so that the metric \eqref{remetric1} of the surface can be written as
\begin{equation}\label{remetric2}
  ds^2=\bigg[1+\left(\frac{dz}{dr}\right)^2\bigg]dr^2+r^2d\phi^2.
\end{equation}
Comparing the equations \eqref{remetric} and \eqref{remetric2}, we can have
\begin{equation}\label{}
  \frac{dz}{dr}=\pm \left(\frac{r}{b(r)}-1\right)^{-\frac{1}{2}}.
\end{equation}
 The geometry of the wormhole solution has least radius at the throat, i.e. $r=b(r)=r_0$, where $r_0$ denotes the radius of the throat of the wormhole. The embedded surface is vertical at the throat, i. e. $\frac{dz}{dr}\to \infty$ at $r=r_0$, and the space is asymptotically flat as $r\to \infty$, i. e. $\frac{dz}{dr}\to 0$ as $r\to\infty$. One also needs to impose the flaring-out condition. The flaring-out condition demands that the inverse of the embedding function $r(z)$ must satisfy $\frac{d^2r}{dz^2}>0$ near or at the throat $r_0$. Now, differentiating
 $\frac{dr}{dz}=\pm\left(\frac{r}{b(r)}-1\right)^{\frac{1}{2}}$ with respect to $z$, we obtain
 \begin{equation}\label{frout}
   \frac{d^2r}{dz^2}=\frac{b(r)-b^{'}(r)r}{2b(r)^2}>0.
 \end{equation}
This flaring-out condition is an important constituent of wormhole physics and plays a major role in the analysis of the violation of the energy conditions. In the light of the above discussion, for a traversable wormhole, the shape function should satisfy the following properties: (i) $\frac{b(r)}{r}<1$ for $r>r_0$,  (ii) $b(r_0)=r_0$ at $r=r_0$, (iii) $\frac{b(r)}{r}\rightarrow 0$ as $r\rightarrow \infty$, (iv) $\frac{b(r)-b'(r)r}{b(r)^2}>0$ for $r>r_0$ and (v) $b'(r)-1< 0$ at $r=r_0$.

\noindent
The energy momentum tensor for the matter source of the wormhole is defined as
\begin{eqnarray}
T_{\mu\nu}&=&\frac{\partial L_m}{\partial g^{\mu\nu}},\nonumber \\
&= &(\rho + p_t)u_\mu u_\nu - p_tg_{\mu\nu}+(p_r-p_t)X_\mu X_\nu,
\end{eqnarray}	
where $\rho$ is the energy density,  $p_t$ and $p_r$ are tangential and radial pressures respectively and $u_{\mu}$ \& $X_\mu$ denote  the four velocity and radial vectors respectively such that
\begin{equation}
u^{\mu}u_\mu=-1 \mbox{ and } X^{\mu}X_\mu=1.
\end{equation}
The effective field equations for the metric \eqref{metric} can be expressed as follows:

\begin{equation}\label{6}
\rho=\frac{Fb'(r)}{r^2}-\Bigg(1-\frac{b(r)}{r}\Bigg)F'\Phi^{'}(r)-H
\end{equation}
\begin{equation}\label{7}
p_r=-\Bigg[\frac{b(r)}{r^3}-2\Bigg(1-\frac{b(r)}{r}\Bigg)\frac{\Phi^{'}(r)}{r}\Bigg]F-\Bigg(1-\frac{b(r)}{r}
\Bigg)\Bigg[F''+\frac{F'(rb'(r)-b(r))}{2r^2\Big(1-\frac{b(r)}{r}\Big)}\Bigg]+H
\end{equation}
\begin{equation}\label{8}
p_t=F\Bigg(1-\frac{b(r)}{r}\Bigg)\Bigg[\Phi^{''}(r)-\frac{(b'(r)r-b(r))}{2r(r-b(r))}\Phi'(r)+\Phi^{'2}+\frac{\Phi^{'}(r)}{r}-
\frac{(b'(r)r-b(r))}{2r^2(r-b(r))}\Bigg]-\frac{F'}{r}\Bigg(1-\frac{b(r)}{r}\Bigg)+H,
\end{equation}
where $F\equiv \frac{df}{dR}$, $R=\Bigg[\frac{4\Phi^{'}(r)}{r}+2\Phi^{''}(r)+2\Phi^{'2}(r)\Bigg]\Bigg(1-\frac{b(r)}{r}\Bigg)-\frac{(b^{'}(r)r-b(r))}{r^2}\Phi^{'}(r)-
\frac{2b^{'}(r)}{r^2}$,

\noindent
 $ H(r)=\frac{1}{4}(FR+\square F+T)$, $\square F=\left(1-\frac{b(r)}{r}\right)\bigg[F^{''}+\frac{rb^{'}(r)-b(r)}{2r^2\left(1-\frac{b(r)}{r}\right)}F^{'}+2\frac{F^{'}}{r}+F^{'}\Phi^{'}(r)\bigg]$ and\\ $T=-\rho+p_r+2p_t$.

\noindent
These are the standard terminologies of the matter threading the wormhole, as a function of the shape function $b(r)$, redshift function $\Phi(r)$ and function $F(r)$.
 One can comprehend the matter content of the wormhole by specifying the above functions. Thus, one can consider a specific choice of shape function to obtain a wormhole solution. Therefore, in this paper a specific form of the shape function $b(r)=\frac{r}{\exp(r-r_0)}$  is considered.

\noindent
Morris and Thorne \cite{morris1} defined the dimensionless function $\xi=\frac{\tau-\rho}{|\rho|}$. However, in this paper, we define the dimensionless function
\begin{equation}\label{xi}
  \xi^{eff}=\frac{-p_r^{eff}-\rho^{eff}}{|\rho^{eff}|}.
\end{equation}
where $\tau= -p_r^{eff}$ and $\rho= \rho^{eff}$.
Now, the equations \eqref{6} and \eqref{7} yields
\begin{equation}\label{xieff}
  \xi^{eff}=\frac{b(r)-rb^{'}(r)}{r|b^{'}(r)|}
\end{equation}
Let us combine the flaring-out condition, given in equation \eqref{frout}, with the equation \eqref{xieff}, the effective exotic function takes the form
\begin{equation}\label{}
  \xi^{eff}=\frac{2b(r)}{r|b^{'}(r)|}\frac{d^2r}{dz^2}
\end{equation}
At the throat, we have the following condition
\begin{equation}\label{}
  \xi(r_0)=\frac{-p_r^{eff}(r_0)-\rho^{eff}(r_0)}{|\rho^{eff}(r_0)|}>0
\end{equation}
Thus, from the above condition it is observed that the radial tension should exceed the total density of mass energy i. e. $-p_r^{eff}(r_0)>\rho^{eff}(r_0)$. We shall call matter with this property, $-p_r^{eff}(r_0)>\rho^{eff}(r_0)>0$, exotic \cite{morris1}. The presence of exotic matter at the throat of the wormhole indicates that the observer who moves through the throat with a radial velocity close to the speed of light will see a negative energy density. Overall, we can say that the field equations plus the absence of a horizon at throat indicates $-p_r^{eff}(r_0)>\rho^{eff}(r_0)$ at the throat. This implies traveler moving through the throat with speed very close to the speed of light can see the negative energy density. This implies the violation of the null, weak, strong and dominant energy conditions at the throat.

\section{Wormhole Solutions} In literature, the cosmologists have studied several models with quantum corrections of Einstein's field equations \cite{Ginzburg, Bunch, Davies}.
Starobinsky followed the same idea and studied the cosmology of a model which in its simplified form is known as Starobinsky model \cite{Star}. He replaced Einstein's action $R$ with $R+\alpha R^2$ and presented a first
compatible model of inflation. In this action, term $R^2$ is a responsible factor for acceleration at high energies in the early stage of the universe. The results obtained from the Planck satellite \cite{Planck} are also consistent with the Starobinsky model. This model has been studied in several aspects \cite{Gorbunov,Kehagias,Ganguly,Paliathanasis,Panotopoulos}. In this paper, a general form of this model $f(R)=R+\alpha R^n$, where $\alpha$ and $n$ are arbitrary constants, is considered. Further, the shape function $b(r)=\frac{r}{\exp(r-r_0)}$ defined by Samanta et al. \cite{godani1}
and the redshift function introduced by Pavlovic and Sossich \cite{Pavlovic} $\Phi(r)=\ln(\frac{r_0}{r}+1)$ is taken into account. The aim is to investigate the  wormhole geometry equipped with less amount of exotic matter near the throat in the setting of $f(R)$ gravity with two redshift functions (i) $\Phi(r) = $ constant and (ii) $\Phi(r)=\ln(\frac{r_0}{r}+1)$. In Section 2, the field equations are derived for arbitrary redshift, shape and $f(R)$ functions. Using forms (i) and (ii) of $\Phi(r)$, the expressions for the energy density ($\rho$), radial pressure ($p_r$), tangential pressure $p_t$ and different combinations of $\rho$, $p_r$ and $p_t$ are determined which are follows:

\subsection{Constant Redshift Function}
\noindent
\textbf{I. $\Phi(r) = $ } constant
\begin{eqnarray}
\rho&=&\frac{\alpha 2^{n-2} e^{r-2 {r_0}} \left(-\frac{(r-1) e^{{r_0}-r}}{r}\right)^{n+1}}{(r-1)^4}\Bigg[ \left(e^{{r_0}} \left(2 n^3 \left(-r^2+r+1\right)^2+n^2 (r (r (r (6-5 r)+12)\right.\right.\nonumber\\
&-& \left.\left.13)-6)+n \left((5 (r-2) r+2) r^2+r+6\right)-2 (r-1)^3 r\right)-2 (n-1) n e^r \left(n \left(-r^2+r+1\right)^2\right.\right.\nonumber\\
&-&\left.\left. r \left(r^3-5 r+4\right)-2\right)\right)\Bigg]-\frac{(r-1) e^{{r_0}-r}}{r^2}\label{d}
\\
p_r&=&\frac{1}{2 r^2}\Bigg[\frac{1}{(r-1)^2}\Bigg(\alpha 2^n r \left(n^2 \left(-2 r^2+2 r+2\right)+n \left(r^3-3\right)-(r-1)^2 r\right) \left(-\frac{(r-1) e^{{r_0}-r}}{r}\right)^n\Bigg)\nonumber\\
&-&\frac{1}{r-1}\Bigg(\alpha 2^{n+1} (n-1) n \left(r^2-r-1\right) \left(-\frac{(r-1) e^{{r_0}-r}}{r}\right)^{n-1}\Bigg)-2 e^{{r_0}-r}\Bigg]\label{pr}
\\
p_t&=&\frac{1}{4 (r-1)^3 r}\Bigg[e^{-r-{r_0}} \left(-\alpha 2^{n+1} n e^{2 r} \left(n^2 \left(-r^2+r+1\right)^2+n \left(-2 r^4+3 r^3+4 r^2-6 r-2\right)\right.\right.\nonumber\\
&+&\left.\left.r^4-r^3-3 r^2+4 r+1\right) \left(-\frac{(r-1) e^{{r_0}-r}}{r}\right)^n+\alpha 2^n \left(2 n^3 \left(-r^2+r+1\right)^2+n^2 \left(-5 r^4\right.\right.\right.\nonumber\\
&+&\left.\left.\left.8 r^3+8 r^2-13 r-4\right)+n \left(5 r^4-11 r^3+2 r^2+6 r+2\right)-2 (r-1)^3 r\right) e^{r+{r_0}}\right.\nonumber\\
 &\times&\left.\left(-\frac{(r-1) e^{{r_0}-r}}{r}\right)^n+2 (r-1)^3 e^{2 {r_0}}\right)\Bigg]\label{pt}
\end{eqnarray}

\noindent
From Equations \eqref{d},\eqref{pr} and \eqref{pt}, the null and dominated energy condition terms are obtained as

\begin{eqnarray}
\rho+p_r&=&\frac{\alpha 2^{n-2} e^{r-2 {r_0}} \left(-\frac{(r-1) e^{{r_0}-r}}{r}\right)^{n+1}}{(r-1)^4}\Bigg[ \left(e^{{r_0}} \left(2 n^3 \left(-r^2+r+1\right)^2+n^2 (r (r (r (6-5 r)+12)\right.\right.\nonumber\\
&-& \left.\left.13)-6)+n \left((5 (r-2) r+2) r^2+r+6\right)-2 (r-1)^3 r\right)-2 (n-1) n e^r \left(n \left(-r^2+r+1\right)^2\right.\right.\nonumber\\
&-&\left.\left. r \left(r^3-5 r+4\right)-2\right)\right)\Bigg]-\frac{(r-1) e^{{r_0}-r}}{r^2}+\frac{1}{2 r^2}\Bigg[\frac{1}{(r-1)^2}\Bigg(\alpha 2^n r \left(n^2 \left(-2 r^2+2 r+2\right)\right.\nonumber\\
&+&\left.n \left(r^3-3\right)-(r-1)^2 r\right) \left(-\frac{(r-1) e^{{r_0}-r}}{r}\right)^n\Bigg)-\frac{1}{r-1}\Bigg(\alpha 2^{n+1} (n-1) n\nonumber
\end{eqnarray}

\begin{eqnarray}
&\times& \left(r^2-r-1\right) \left(-\frac{(r-1) e^{{r_0}-r}}{r}\right)^{n-1}\Bigg)-2 e^{{r_0}-r}\Bigg]
\end{eqnarray}
\begin{eqnarray}
\rho+p_t&=&\frac{\alpha 2^{n-2} e^{r-2 {r_0}} \left(-\frac{(r-1) e^{{r_0}-r}}{r}\right)^{n+1}}{(r-1)^4}\Bigg[ \left(e^{{r_0}} \left(2 n^3 \left(-r^2+r+1\right)^2+n^2 (r (r (r (6-5 r)+12)\right.\right.\nonumber\\
&-& \left.\left.13)-6)+n \left((5 (r-2) r+2) r^2+r+6\right)-2 (r-1)^3 r\right)-2 (n-1) n e^r \left(n \left(-r^2+r+1\right)^2\right.\right.\nonumber\\
&-&\left.\left. r \left(r^3-5 r+4\right)-2\right)\right)\Bigg]-\frac{(r-1) e^{{r_0}-r}}{r^2}+\frac{1}{4 (r-1)^3 r}\Bigg[e^{-r-{r_0}} \left(-\alpha 2^{n+1} n e^{2 r} \left(n^2\right.\right.\nonumber\\
&\times&\left.\left. \left(-r^2+r+1\right)^2+n \left(-2 r^4+3 r^3+4 r^2-6 r-2\right)+r^4-r^3-3 r^2+4 r+1\right)\right.\nonumber\\
&\times&\left. \left(-\frac{(r-1) e^{{r_0}-r}}{r}\right)^n+\alpha 2^n \left(2 n^3 \left(-r^2+r+1\right)^2+n^2 \left(-5 r^4+8 r^3+8 r^2-13 r-4\right)\right.\right.\nonumber\\
&+&\left.\left.n \left(5 r^4-11 r^3+2 r^2+6 r+2\right)-2 (r-1)^3 r\right) e^{r+{r_0}}\left(-\frac{(r-1) e^{{r_0}-r}}{r}\right)^n
\right.\nonumber\\
 &+&\left.2 (r-1)^3 e^{2 {r_0}}\right)\Bigg]
 \end{eqnarray}

\begin{eqnarray}
\rho - |p_r|&=&\frac{\alpha 2^{n-2} e^{r-2 {r_0}} \left(-\frac{(r-1) e^{{r_0}-r}}{r}\right)^{n+1}}{(r-1)^4}\Bigg[ \left(e^{{r_0}} \left(2 n^3 \left(-r^2+r+1\right)^2+n^2 (r (r (r (6-5 r)+12)\right.\right.\nonumber\\
&-& \left.\left.13)-6)+n \left((5 (r-2) r+2) r^2+r+6\right)-2 (r-1)^3 r\right)-2 (n-1) n e^r \left(n \left(-r^2+r+1\right)^2\right.\right.\nonumber\\
&-&\left.\left. r \left(r^3-5 r+4\right)-2\right)\right)\Bigg]-\frac{(r-1) e^{{r_0}-r}}{r^2}-\Bigg|\frac{1}{2 r^2}\Bigg[\frac{1}{(r-1)^2}\Bigg(\alpha 2^n r \left(n^2 \left(-2 r^2+2 r+2\right)\right.\nonumber\\
&+&\left.n \left(r^3-3\right)-(r-1)^2 r\right) \left(-\frac{(r-1) e^{{r_0}-r}}{r}\right)^n\Bigg)-\frac{1}{r-1}\Bigg(\alpha 2^{n+1} (n-1) n\nonumber\\
&\times& \left(r^2-r-1\right) \left(-\frac{(r-1) e^{{r_0}-r}}{r}\right)^{n-1}\Bigg)-2 e^{{r_0}-r}\Bigg]\Bigg|
\end{eqnarray}
\begin{eqnarray}
\rho-|p_t|&=&\frac{\alpha 2^{n-2} e^{r-2 {r_0}} \left(-\frac{(r-1) e^{{r_0}-r}}{r}\right)^{n+1}}{(r-1)^4}\Bigg[ \left(e^{{r_0}} \left(2 n^3 \left(-r^2+r+1\right)^2+n^2 (r (r (r (6-5 r)+12)\right.\right.\nonumber\\
&-& \left.\left.13)-6)+n \left((5 (r-2) r+2) r^2+r+6\right)-2 (r-1)^3 r\right)-2 (n-1) n e^r \left(n \left(-r^2+r+1\right)^2\right.\right.\nonumber
\end{eqnarray}

\begin{eqnarray}
&-&\left.\left. r \left(r^3-5 r+4\right)-2\right)\right)\Bigg]-\frac{(r-1) e^{{r_0}-r}}{r^2}-\Bigg|\frac{1}{4 (r-1)^3 r}\Bigg[e^{-r-{r_0}} \left(-\alpha 2^{n+1} n e^{2 r} \left(n^2\right.\right.\nonumber\\
&\times&\left.\left. \left(-r^2+r+1\right)^2+n \left(-2 r^4+3 r^3+4 r^2-6 r-2\right)+r^4-r^3-3 r^2+4 r+1\right)\right.\nonumber\\
&\times&\left. \left(-\frac{(r-1) e^{{r_0}-r}}{r}\right)^n+\alpha 2^n \left(2 n^3 \left(-r^2+r+1\right)^2+n^2 \left(-5 r^4+8 r^3+8 r^2-13 r-4\right)\right.\right.\nonumber\\
&+&\left.\left.n \left(5 r^4-11 r^3+2 r^2+6 r+2\right)-2 (r-1)^3 r\right) e^{r+{r_0}}\left(-\frac{(r-1) e^{{r_0}-r}}{r}\right)^n
\right.\nonumber\\
 &+&\left.2 (r-1)^3 e^{2 {r_0}}\right)\Bigg]\Bigg|\end{eqnarray}

\begin{eqnarray}
p_t-p_r&=&\frac{1}{4 (r-1)^3 r}\Bigg[e^{-r-{r_0}} \left(-\alpha 2^{n+1} n e^{2 r} \left(n^2 \left(-r^2+r+1\right)^2+n \left(-2 r^4+3 r^3+4 r^2-6 r-2\right)\right.\right.\nonumber\\
&+&\left.\left.r^4-r^3-3 r^2+4 r+1\right) \left(-\frac{(r-1) e^{{r_0}-r}}{r}\right)^n+\alpha 2^n \left(2 n^3 \left(-r^2+r+1\right)^2+n^2 \left(-5 r^4\right.\right.\right.\nonumber\\
&+&\left.\left.\left.8 r^3+8 r^2-13 r-4\right)+n \left(5 r^4-11 r^3+2 r^2+6 r+2\right)-2 (r-1)^3 r\right) e^{r+{r_0}}\right.\nonumber\\
 &\times&\left.\left(-\frac{(r-1) e^{{r_0}-r}}{r}\right)^n+2 (r-1)^3 e^{2 {r_0}}\right)\Bigg]-\frac{1}{2 r^2}\Bigg[\frac{1}{(r-1)^2}\Bigg(\alpha 2^n r \left(n^2 \left(-2 r^2+2 r+2\right)\right.\nonumber\\
&+&\left.n \left(r^3-3\right)-(r-1)^2 r\right) \left(-\frac{(r-1) e^{{r_0}-r}}{r}\right)^n\Bigg)-\frac{1}{r-1}\Bigg(\alpha 2^{n+1} (n-1) n \left(r^2-r-1\right)\nonumber\\
&\times& \left(-\frac{(r-1) e^{{r_0}-r}}{r}\right)^{n-1}\Bigg)-2 e^{{r_0}-r}\Bigg]
\end{eqnarray}

\begin{eqnarray}
\frac{p_r}{\rho}&=&\frac{1}{2 r^2}\Bigg[\frac{1}{(r-1)^2}\Bigg(\alpha 2^n r \left(n^2 \left(-2 r^2+2 r+2\right)+n \left(r^3-3\right)-(r-1)^2 r\right) \left(-\frac{(r-1) e^{{r_0}-r}}{r}\right)^n\Bigg)\nonumber\\
&-&\frac{1}{r-1}\Bigg(\alpha 2^{n+1} (n-1) n \left(r^2-r-1\right) \left(-\frac{(r-1) e^{{r_0}-r}}{r}\right)^{n-1}\Bigg)-2 e^{{r_0}-r}\Bigg]\nonumber\\
&\div&\Bigg[\frac{\alpha 2^{n-2} e^{r-2 {r_0}} \left(-\frac{(r-1) e^{{r_0}-r}}{r}\right)^{n+1}}{(r-1)^4}\Bigg[ \left(e^{{r_0}} \left(2 n^3 \left(-r^2+r+1\right)^2+n^2 (r (r (r (6-5 r)+12)\right.\right.\nonumber\\
&-& \left.\left.13)-6)+n \left((5 (r-2) r+2) r^2+r+6\right)-2 (r-1)^3 r\right)-2 (n-1) n e^r \left(n \left(-r^2+r+1\right)^2\right.\right.\nonumber\\
&-&\left.\left. r \left(r^3-5 r+4\right)-2\right)\right)\Bigg]-\frac{(r-1) e^{{r_0}-r}}{r^2}\Bigg]
\end{eqnarray}

\subsection{Variable Redshift Function}
\noindent
\textbf{II. $\Phi(r)=\ln(\frac{r_0}{r}+1)$}
\begin{eqnarray}
\rho&=&\frac{1}{\left(2 r^2+2 (r_0-1) r-3 r_0\right)^3}+\frac{1}{2}\Bigg[\alpha e^{r-2 r_0} \left(-e^r+e^{r_0}\right) (n-1) n r_0 \left(2 r^4+(4 r_0-2) r^3+\left(2 r_0^2\right.\right.\nonumber\\
&-&\left.\left.5 r_0-2\right) r^2-3 r_0 (r_0+2) r-3 r_0^2\right) \left(\frac{e^{r_0-r} \left(2 r^2+2 (r_0-1) r-3 r_0\right)}{r (r+r_0)}\right)^{n+1}\Bigg] \nonumber\\
&\times&\left(\frac{e^{r_0-r} \left(-2 r^2-2 (r_0-1) r+3 r_0\right)}{r (r+r_0)}-\alpha n \left(\frac{e^{r_0-r} \left(2 r^2+2 (r_0-1) r-3 r_0\right)}{r (r+r_0)}\right)^n\right)+\frac{1}{2} \left(\alpha\right. \nonumber\\
&\times&\left.\left(\frac{e^{r_0-r} \left(2 r^2+2 (r_0-1) r-3 r_0\right)}{r (r+r_0)}\right)^n+\frac{e^{r_0-r} \left(2 r^2+2 (r_0-1) r-3 r_0\right)}{r (r+r_0)}\right)+\alpha e^{r-2 r_0} (n-1) n\nonumber\\
&\times&\frac{1}{2 \left(2 r^2+2 (r_0-1) r-3 r_0\right)^4}\Bigg[ \left(\frac{e^{r_0-r} \left(2 r^2+2 (r_0-1) r-3 r_0\right)}{r (r+r_0)}\right)^{n+1} \left(e^{r_0} \left(4 (2 n-3) r^8\right.\right.\nonumber\\
&+&\left.\left.8 (-6 r_0+n (4 r_0-2)+1) r^7+\left(-72 r_0^2+44 r_0+8 n \left(6 r_0^2-9 r_0-1\right)+40\right) r^6+4 \left(-12 r_0^3+21 r_0^2\right.\right.\right.\nonumber\\
&+&\left.\left.\left.46 r_0+2 n \left(4 r_0^3-15 r_0^2-5 r_0+2\right)-9\right) r^5+\left(-12 r_0^4+68 r_0^3+299 r_0^2-198 r_0+n \left(8 r_0^4-88 r_0^3\right.\right.\right.\right.\nonumber\\
&-&\left.\left.\left.\left.62 r_0^2+88 r_0+8\right)-16\right) r^4+2 r_0 \left(10 r_0^3+103 r_0^2-183 r_0-6 n \left(2 r_0^3+3 r_0^2-14 r_0-4\right)-44\right) r^3\right.\right.\nonumber\\
&-&\left.\left.3 r_0^2 \left(-17 r_0^2+95 r_0+2 n \left(r_0^2-22 r_0-16\right)+60\right) r^2+9 r_0^3 (-9 r_0+4 n (r_0+2)-16) r+18 (n-2) \right.\right.\nonumber\\
&\times&\left.\left.r_0^4\right)-2 e^r \left(4 (n-1) r^8+8 (n (2 r_0-1)-2 r_0) r^7+4 \left(-6 r_0^2+r_0+n \left(6 r_0^2-9 r_0-1\right)+5\right) r^6 \right.\right.\nonumber\\
&+&\left.\left.4 \left(-4 r_0^3+3 r_0^2+23 r_0+n \left(4 r_0^3-15 r_0^2-5 r_0+2\right)-4\right) r^5+\left(-4 r_0^4+12 r_0^3+151 r_0^2-88 r_0\right.\right.\right.\nonumber\\
&+&\left.\left.\left.n \left(4 r_0^4-44 r_0^3-31 r_0^2+44 r_0+4\right)-8\right) r^4-2 r_0 \left(-2 r_0^3-53 r_0^2+81 r_0+3 n \left(2 r_0^3+3 r_0^2-14 r_0\right.\right.\right.\right.\nonumber\\
&-&\left.\left.\left.\left.4\right)+22\right) r^3-3 r_0^2 \left(-9 r_0^2+42 r_0+n \left(r_0^2-22 r_0-16\right)+30\right) r^2+18 (n-2) r_0^3 (r_0+2) r\right.\right.\nonumber\\
&+&\left.\left.9 (n-2) r_0^4\right)\right)\Bigg]-\frac{1e^{r_0-r} (r-1)}{r^2} \left(\alpha n \left(\frac{e^{r_0-r} \left(2 r^2+2 (r_0-1) r-3 r_0\right)}{r (r+r_0)}\right)^{n-1}+1\right)
\end{eqnarray}

\begin{eqnarray}
p_r&=&\frac{1}{2 r^2 \left(2 r^2+2 (r_0-1) r-3 r_0\right)^2 (r+r_0)}\Bigg[4 \alpha (n-1) r^7 \left(\frac{e^{r_0-r} \left(2 r^2+2 (r_0-1) r-3 r_0\right)}{r (r+r_0)}\right)^n\nonumber\\
&+&4 \alpha (n-1) r^6 (2 n+3 r_0-2) \left(\frac{e^{r_0-r} \left(2 r^2+2 (r_0-1) r-3 r_0\right)}{r (r+r_0)}\right)^n+4 \alpha r^5 \left((6 r_0-2) n^2\right.\nonumber\\
&+&\left.\left(3 r_0^2-13 r_0+2\right) n-3 r_0^2+7 r_0-1\right) \left(\frac{e^{r_0-r} \left(2 r^2+2 (r_0-1) r-3 r_0\right)}{r (r+r_0)}\right)^n-4 \alpha e^{r-r_0} n r \nonumber\\
&\times&(r+r_0) \left(2 (n-1) r^4+2 (n-1) (2 r_0-1) r^3+\left(-2 r_0^2+7 r_0+n \left(2 r_0^2-5 r_0-2\right)+2\right) r^2\right.\nonumber\\
&+&\left.r_0 (5 r_0-3 n (r_0+2)+4) r-3 n r_0^2\right) \left(\frac{e^{r_0-r} \left(2 r^2+2 (r_0-1) r-3 r_0\right)}{r (r+r_0)}\right)^n-36 r_0^3-2 e^{r_0-r} \nonumber\\
&\times&(r-r_0) \left(2 r^2+2 (r_0-1) r-3 r_0\right)^2-6 r r_0^2 \left(r_0 \left(\alpha n (2 n-1) \left(\frac{e^{r_0-r} \left(2 r^2+2 (r_0-1) r-3 r_0\right)}{r (r+r_0)}\right)^n\right.\right.\nonumber\\
&-&\left.\left.8\right)+8\right)+4 r^4 \left(\alpha (n-1) r_0^3 \left(\frac{e^{r_0-r} \left(2 r^2+2 (r_0-1) r-3 r_0\right)}{r (r+r_0)}\right)^n+2 \alpha \left(3 n^2-7 n+4\right) r_0^2\right.\nonumber
\end{eqnarray}

\begin{eqnarray}
&\times&\left.\left(\frac{e^{r_0-r} \left(2 r^2+2 (r_0-1) r-3 r_0\right)}{r (r+r_0)}\right)^n-\alpha n (2 n-3) \left(\frac{e^{r_0-r} \left(2 r^2+2 (r_0-1) r-3 r_0\right)}{r (r+r_0)}\right)^n\right.\nonumber\\
&-&\left.r_0 \left(\alpha \left(7 n^2-10 n+4\right) \left(\frac{e^{r_0-r} \left(2 r^2+2 (r_0-1) r-3 r_0\right)}{r (r+r_0)}\right)^n+4\right)\right)-r^2 r_0 \left(\left(\alpha \left(12 n^2-25 n+9\right)\right.\right.\nonumber\\
&\times&\left.\left. \left(\frac{e^{r_0-r} \left(2 r^2+2 (r_0-1) r-3 r_0\right)}{r (r+r_0)}\right)^n+16\right) r_0^2+4 \left(\alpha n (9 n-8) \left(\frac{e^{r_0-r} \left(2 r^2+2 (r_0-1) r-3 r_0\right)}{r (r+r_0)}\right)^n\right.\right.\nonumber\\
&-&\left.\left.20\right) r_0+16\right)+r^3 r_0 \left(4 \alpha \left(2 n^2-5 n+3\right) r_0^2 \left(\frac{e^{r_0-r} \left(2 r^2+2 (r_0-1) r-3 r_0\right)}{r (r+r_0)}\right)^n-2 \alpha n (16 n-19)\right.\nonumber\\
&\times& \left.\left(\frac{e^{r_0-r} \left(2 r^2+2 (r_0-1) r-3 r_0\right)}{r (r+r_0)}\right)^n-r_0 \left(\alpha \left(32 n^2-57 n+21\right) \left(\frac{e^{r_0-r} \left(2 r^2+2 (r_0-1) r-3 r_0\right)}{r (r+r_0)}\right)^n\right.\right.\nonumber\\
&+&\left.\left.32\right)32\right)\Bigg]
\end{eqnarray}

\begin{eqnarray}
p_t&=& \frac{1}{2 r^2 (r+r_0)}\Bigg[\frac{1}{r (r+r_0)}\Bigg(2 \alpha e^{r_0-2 r} \left(e^r-e^{r_0}\right) (n-1) n \left(2 r^4+(4 r_0-2) r^3+\left(2 r_0^2-5 r_0-2\right) r^2\right.\nonumber\\
&-&\left.3 r_0 (r_0+2) r-3 r_0^2\right) \left(\frac{e^{r_0-r}}{r (r+r_0)} \left(2 r^2+2 (r_0-1) r-3 r_0\right)\right)^{n-2}\Bigg)+r \left(\alpha n r (r+r_0) \left(\frac{e^{r_0-r}}{r (r+r_0)}\right.\right.\nonumber\\
&\times&\left.\left.\left(2 r^2+2 (r_0-1) r-3 r_0\right)\right)^n+e^{r_0-r} \left(2 r^2+2 (r_0-1) r-3 r_0\right)\right)+\frac{1}{2 r^2+2 (r_0-1) r-3 r_0}\Bigg[e^{-r-r_0}\nonumber\\
&\times& \left(2 e^r r_0+e^{r_0} \left(r^2-2 r_0\right)\right) \left(\alpha e^r n r (r+r_0) \left(\frac{e^{r_0-r} \left(2 r^2+2 (r_0-1) r-3 r_0\right)}{r (r+r_0)}\right)^n+e^{r_0} \left(2 r^2+2 (r_0\right.\right.\nonumber\\
&-&\left.\left.1) r-3 r_0\right)\right)\Bigg]+r \left(e^{r_0-r} \left(-2 r^2-2 (r_0-1) r+3 r_0\right)-\alpha r \left(\frac{e^{r_0-r} \left(2 r^2+2 (r_0-1) r-3 r_0\right)}{r (r+r_0)}\right)^n\right.\nonumber\\
&\times&\left.(r+r_0)\right)+\frac{1}{\left(2 r^2+2 (r_0-1) r-3 r_0\right)^3}\Bigg(\alpha e^{-r_0} (n-1) n r \left(\frac{e^{r_0-r} \left(2 r^2+2 (r_0-1) r-3 r_0\right)}{r (r+r_0)}\right)^n\nonumber\\
&\times& \left(2 e^r \left(4 (n-1) r^8+8 (n (2 r_0-1)-2 r_0) r^7+4 \left(-6 r_0^2+r_0+n \left(6 r_0^2-9 r_0-1\right)+5\right) r^6\right.\right.\nonumber\\
&+&\left.\left.4 \left(-4 r_0^3+3 r_0^2+23 r_0+n \left(4 r_0^3-15 r_0^2-5 r_0+2\right)-4\right) r^5+\left(-4 r_0^4+12 r_0^3+151 r_0^2-88 r_0\right.\right.\right.\nonumber\\
&+&\left.\left.\left.n \left(4 r_0^4-44 r_0^3-31 r_0^2+44 r_0+4\right)-8\right) r^4-2 r_0 \left(-2 r_0^3-53 r_0^2+81 r_0+3 n \left(2 r_0^3+3 r_0^2-14 r_0\right.\right.\right.\right.\nonumber\\
&-&\left.\left.\left.\left.4\right)+22\right) r^3-3 r_0^2 \left(-9 r_0^2+42 r_0+n \left(r_0^2-22 r_0-16\right)+30\right) r^2+18 (n-2) r_0^3 (r_0+2) r\right.\right.\nonumber\\
&+&\left.\left.9 (n-2) r_0^4\right)+e^{r_0} \left((12-8 n) r^8-8 (-6 r_0+n (4 r_0-2)+1) r^7+\left(72 r_0^2-44 r_0+n \left(-48 r_0^2\right.\right.\right.\right.\nonumber\\
&+&\left.\left.\left.\left.72 r_0+8\right)-40\right) r^6-4 \left(-12 r_0^3+21 r_0^2+46 r_0+2 n \left(4 r_0^3-15 r_0^2-5 r_0+2\right)-9\right) r^5+\left(12 r_0^4\right.\right.\right.\nonumber\\
&-&\left.\left.\left.68 r_0^3-299 r_0^2+198 r_0+n \left(-8 r_0^4+88 r_0^3+62 r_0^2-88 r_0-8\right)+16\right) r^4+2 r_0 \left(-10 r_0^3-103 r_0^2\right.\right.\right.\nonumber\\
&+&\left.\left.\left.183 r_0+6 n \left(2 r_0^3+3 r_0^2-14 r_0-4\right)+44\right) r^3+3 r_0^2 \left(-17 r_0^2+95 r_0+2 n \left(r_0^2-22 r_0-16\right)\right.\right.\right.\nonumber\\
&+&\left.\left.\left.60\right) r^2-9 r_0^3 (-9 r_0+4 n (r_0+2)-16) r-18 (n-2) r_0^4\right)\right)\Bigg)\Bigg]
\end{eqnarray}
Similar to $\phi(r) = $ constant, the expressions for $\rho+p_r$, $\rho+p_t$, $\rho+p_r+2p_t$, $\rho-|p_r|$, $\rho-|p_t|$, $p_t-p_r$ and $\frac{p_r}{\rho}$ can be determined. Since the expressions for $\rho$, $p_r$ and $p_t$ are too large, therefore we are not mentioning these combinations here.

\noindent
The geometric nature of wormholes can be determined using the anisotropy parameter which is defined as $\triangle=p_t-p_r$. For  $\triangle>0$, the geometry is said to be repulsive; for $\triangle<0$, the geometry is said to be attractive and for $\triangle=0$, the geometry is called isotropic.
The equation of state parameter is defined as $\omega=\frac{p_r}{\rho}$. Its value determines the type of the fluid present in the wormhole structure.

\section{Energy Conditions}
In the literature \cite{morris1}, it is	 pointed out that not only a throat of the spherically static wormhole threaded by exotic matter, but this is also true for any traversable, non-static and non-spherical wormhole. The main reason is that the bundles of null geodesics that enter the wormhole at one end (mouth) and arise from the other must have cross-sectional areas that initially decrease and then increase. The translation from shrinking to growing can only be formed by gravitational repulsion of matter through which the light rays pass. So, negative energy density is required for this repulsion \cite{morris1, Visser}.
In the 1960s and an early 1970s, most physicists claim that no observer should ever be able to measure a negative energy density. This claim brings the name weak energy condition, and when this improved by some additional limitations, it is called the dominant energy condition or the strong energy condition. These energy conditions are allowed to violate, for the matter with the property $-p_r>\rho$.  And, these are key foundations for a number of important theorems, for example: the positive mass theorem, which says that objects made of matter can never repel other bodies gravitationally, provided it satisfies the dominant energy condition \cite{Schoen, Schoen1, Witten, Parker, Gibbons, Moreschi}. A variety of theorems that forecast that if one or more of the energy conditions are satisfied, then space time singularities will be formed in cosmological situations and in gravitational collapse \cite{Hawking}, and the second law of black hole mechanics, which says that if  stress energy near a black hole horizon satisfies the strong energy condition, then the horizon’s surface area can never decrease \cite{Hawking1}.

\noindent
The energy momentum tensor at every point $x\in \mathbb{M}$ must follow the inequality $T_{\mu\nu}W^{\mu}W^{\nu}\ge 0$ for any timelike vector
$W\in \mathbb{T}_x$, where $\mathbb{M}$ is the 4-dimensional space-time and $\mathbb{T}_x$ is the tangent space at $x\in \mathbb{M}$. And, this will be true for any null vector $W\in \mathbb{T}_x$.
An observer whose world line at $x$ has unit tangent vector $U$, the
local energy density seems to be $T_{\mu\nu}U^{\mu}U^{\nu}$. Thus, this supposition is
corresponding to saying that the energy density as measured by any
observer is non-negative. That is $\mbox{NEC}\Leftrightarrow ~~ \rho+p_i\ge 0, ~\forall i$. For our model, the null energy condition (NEC) is said to be satisfied, if $\rho + p_r\geq0$ and $\rho+p_t\geq 0$ for all $r>0$.
The Week Energy Condition (WEC) is defined as $\mbox{WEC} \Leftrightarrow T_{\mu\nu}W^{\mu}W^{\nu}\ge 0$, where $W^{\mu}$ is any time like vector. As it is true for any timelike vector, so it will also suggest the Null Energy Condition (NEC). The physical significance of this condition is that it claims the local energy density must be positive as measured by any timelike observer. That is $\mbox{WEC}\Leftrightarrow \rho\ge 0, ~~ \mbox{and} ~~ \rho+p_i\ge 0, ~\forall i$. For our model, the week energy condition (WEC) is said to be satisfied, if $\rho\geq 0$, $\rho + p_r\geq0$ and $\rho+p_t\geq 0$ for all $r>0$.
For any timelike vector $W^{\mu}$, the Strong Energy Condition (SEC) is defined as $\mbox{SEC}\Leftrightarrow \left(T_{\mu\nu}-\frac{T}{2}g_{\mu\nu}\right)W^{\mu}W^{\nu}\ge 0$, where $T$ is the trace of the stress energy tensor, $T=T_{\mu\nu}g^{\mu\nu}$. The SEC also suggest the NEC, but it does not imply, in general, the WEC. Precisely, $\mbox{SEC}\Leftrightarrow \rho+p_i\ge 0, \mbox{and } \rho+\sum p_i\ge 0, \forall i$. For our model, the strong energy condition (SEC) is said to be satisfied, if $\rho + p_r\geq0$, $\rho+p_t\geq 0$ and $\rho+p_r+2p_t\geq 0$ for all $r>0$. For any timelike vector $W^{\mu}$, the Dominant Energy Condition (DEC) is defined as $\mbox{DEC}\Leftrightarrow T_{\mu\nu}W^{\mu}W^{\nu}\ge 0$,
and $T_{\mu\nu}W^{\mu}$ is not spacelike. The physical significance of this energy condition says that the energy density will be always positive locally, and that the energy flux is timelike or null. The dominant energy condition (DEC) is said to be satisfied if $\rho-|p_r|\geq 0$ and $\rho-|p_t|\geq 0$ for all $r>0$. The DEC implies the WEC, and thus also the NEC, however, it does not necessarily imply the SEC. Precisely, we can write $\mbox{DEC}\Leftrightarrow \rho\ge 0, \mbox{and } p_i\in [-\rho, +\rho], \forall i$.

\noindent
In modified gravity the gravitational field equations can be rewritten
as an effective Einstein equation, given by $G_{\mu\nu}=\kappa^2T_{\mu\nu}^{eff}$, where $T_{\mu\nu}^{eff}$ is an stress energy tensor containing the stress energy tensor and curvature, arising from the specific modified gravity considered\cite{Harko}. Hence the generalized NEC for the modified gravity is defined as $T_{\mu\nu}^{eff}W^{\mu}W^{\nu}\ge 0$. The violation of the NEC is necessary, for the existence of wormhole solution. Therefore, in modified gravity, the violation of the generalized NEC is necessary, for the existence of wormhole solution. Hence, $T_{\mu\nu}^{eff}W^{\mu}W^{\nu}< 0$ is required. This may reduce the violation of the NEC in classical general relativity, i. e. $T_{\mu\nu}W^{\mu}W^{\nu}<0$. In order to ensure the flaring-out condition, the generalized NEC is required to be violated, i. e.   $T_{\mu\nu}^{eff}W^{\mu}W^{\nu}< 0$ . Moreover, in modified gravity one may impose some constraints, such that the matter stress energy tensor satisfies the standard NEC, i. e.  $T_{\mu\nu}W^{\mu}W^{\nu}\ge 0$, while the respective generalized NEC will be violated.
From equations \eqref{6} and \eqref{7}, it is observed that $\rho^{eff}(r_0)+p_r^{eff}(r_0)=-\frac{1}{r_0}$, this indicates that the generalized NEC does not satisfy at the throat $r_0$ of the wormhole, which supports the existence of traversable wormhole in modified gravity.

\section{Results \& Discussion}
In the exploration of wormhole geometries, the modified theories have contributed significantly. The modified $f(R)$ theory is one which has been used by several cosmologists to study the wormhole geometry. The wormhole metric is defined in terms of shape and redshift functions. We have considered   variable redshift function $\phi(r) = \log(\frac{r_0}{r}+1)$, proposed by  P. Pavlovic and M. Sossich \cite{Pavlovic},  as well as  constant redshift function $\phi(r)= $ constant with the shape function $b(r)=\frac{r}{\exp(r-r_0)}$, introduced by Samanta et al. \cite{godani1}. The  energy conditions are investigated to obtain the wormhole geometries in the framework of $f(R)=R+\alpha R^n$ gravity, where $\alpha$ and $n$ are arbitrary constants. Since $R$ is a function of $r$, $f(R)$ depends on $r$, $\alpha$ and $n$.
For different possible values of these parameters, the validity of  energy conditions and nature of anisotropy and equation of state parameters, calculated in Section 3, are analyzed. Further, the  spherical regions obeying the energy conditions are also determined.  \\


\noindent
\textbf{I: $\phi(r)= constant$}\\
\noindent
\textbf{Case I(a): $\alpha=0$}\\
In this case, $f(R)=R$. The results are summarized in Table 1. The energy density is  positive only for $0<r<1$. The first NEC term $\rho + p_r$ is negative for $r\in(0,\infty)-{1}$ and indeterminate for $r=1$. This shows that NEC and hence WEC are not satisfied everywhere. The first DEC term $\rho-|p_r|< 0$  for all $r\in(0,\infty)-\{1\}$. This shows the violation of DEC throughout. Thus, this case is not of interest.   \\

\noindent
\textbf{Case I(b): $n=0$}\\
In this case, $f(R)=R+\alpha$. The results are summarized in Table 2. It is clear from Table 2 that all the energy conditions are also violated here like case I(a). So, this case does not give favourable results.   \\

\noindent
\textbf{Case I(c): $n>0$}\\
If $n$ is not an integer, then the energy density and all energy condition terms have either negative, imaginary or indeterminate values for $r\in (0,\infty)$. If $n$ is a positive integer, then we have different results for $n=1,2$ and $>2$.
For $n>2$, we have taken $n=6$.
In Tables 3-5, the results are summarized for $n=1,2$ and 6 respectively with the variation of $\alpha$ and $r$.\\

\noindent
When $n=1$  and $\alpha>0$, the energy density is positive only for $r\in(0,1)$ and all energy conditions are violated. For When $n=1$  and $\alpha<0$,  NEC, WEC and DEC are satisfied for $r\in(2,\infty)$. In this region, the matter filled is ordinary with attractive geometry.  Thus, energy conditions are violated near the throat of wormhole and satisfied away from it.\\

\noindent
When $n=2$, $f(R)=R+\alpha R^2$. The results are specified in Table 4. For $\alpha>0$, there is no spherical region obeying the energy conditions. For $\alpha<0$, the energy density is positive for $r\in [0.4,2.7]$ and NEC is satisfied for $r\in(1.5,2.1)$. Consequently, WEC is valid for  $r\in(1.5,2.1)$. Further, SEC is satisfied for $r\in(1.5,2.1)$ and DEC is satisfied for $r\in(1.5,2.1]$. Thus, all energy conditions are valid for $r\in(1.5,2.1)$. In this range, the geometry is filled with non-phantom fluid having attractive geometry. Thus, we have desired results but in a very small region.\\

\noindent
When $n=6$, $f(R)=R+\alpha R^6$, the results obtained are summarized in Table 5. For $\alpha>0$, all the energy conditions are found to be violated. For $\alpha<0$, the energy density is positive for $r\in (0.1,1)$ and NEC is satisfied for $r\in(0.8,1)$. Thus, WEC is satisfied for $r\in(0.8,1)$. In this region, the geometry is repulsive and matter content is ordinary. Both SEC and DEC are violated throughout.  So we do not have good results for $n>2$.
\\

\noindent
\textbf{Case I(d): $n<0$}
We have found  different results for $n=-2$,  $n\neq -2$ but a negative integer and $n$ not an integer. For $n\neq -2$ but a negative integer, we have taken $n=-1$. Thus, we have summarized the results, in particular, for $n=-0.1, -1$ and -2 in Tables 6-8 respectively.\\

\noindent
For $n=-0.1$, $f(R)=R+\alpha R^{-0.1}$. The results obtained are mentioned in Table 6. If $\alpha>0$, the energy density is positive for $r\in(0,1)$ and NEC is satisfied for $r\in[0.4,1)$. Consequently, WEC is valid for $r\in[0.4,1)$. For this range of $r$, the matter content is non-phantom with attractive geometry. SEC and DEC are violated everywhere. If $\alpha<0$, the energy conditions are dissatisfied everywhere.\\

\noindent
For $n=-1$, $f(R)=R+ \frac{\alpha}{R}$. In Table 7,  the results are declared for $\alpha>0$ and $\alpha<0$.  If $\alpha>0$, we have $\rho>0$ for $r\in (0,\infty)-\{1\}$ and NEC validates for  $r\in (0.4,1)\cup(1.6,\infty)$. Thus, WEC is also satisfied for $r\in (0.4,1)\cup(1.6,\infty)$. SEC is satisfied nowhere and DEC is satisfied for $r\in (0.3,1)\cup(1.6,\infty)$. Hence, NEC, WEC and DEC are satisfied for $r\in (0.4,1)\cup(1.6,\infty)$ with non-phantom fluid having attractive geometry. If $\alpha<0$, energy conditions are violated everywhere. This depicts the presence of exotic matter with repulsive geometric configuration near the throat for $r\in(0,0.4)$.  So, we have desirable results in this subcase. These results are  plotted in Figures 1(a)-1(g). \\

\noindent
For $n=-2$, $f(R)=R+ \frac{\alpha}{R^2}$. In Table 8,  the results are specified for $\alpha>0$ and $\alpha<0$. When $\alpha>0$, $\rho>0$ for $r\in(0,1)$ and NEC, WEC, SEC and DEC are  satisfied for $r\in(0.3,1)$. In this range of $r$, the geometry is filled with non-phantom fluid having attractive geometry.  Thus, the results are favourable but in a small spherical region. When $\alpha<0$,  $\rho>0$ for $r\in(1,\infty)$. The first NEC term is positive for $r>1$ and second NEC term is positive for $r\in(0,0.4)\cup(1.6,\infty)$. Thus, NEC and WEC are obeyed for  $r>1.6$. The SEC term $\rho+p_r+2p_t$ is positive for $r\in(0,1)$ only, therefore SEC is disobeyed everywhere. Futhere, the first and second DEC terms are positive for $r>1$ and $r>1.6$ respectively. Thus, DEC is satisfied for  $r>1.6$. The geometric structure is filled with ordinary or non-phantom fluid with repulsive geometry near the throat for $r<1$ and attractive geometry for $r>1$.
\\


\noindent
\textbf{II: $\phi(r)=\ln(\frac{r_0}{r}+1)$}\\
\noindent
\textbf{Case II(a):} $\alpha=0$\\
In this case, the energy density is negative  and all  energy conditions are invalid everywhere. \\

\noindent
\textbf{Case II(b):} $n=0$\\
The results are similar to Case II(a) for every value of $r$ and $\alpha$. \\

\noindent
\textbf{Case II(c):}  $n>0$\\
The parameter $\alpha$ can be positive or negative. When $\alpha>0$, the results are similar to Case II(a) everywhere. When $\alpha<0$, the energy density is observed to be positive for $r>1$ with $\alpha\leq -1$ and $n=1$ and negative otherwise. For $\alpha\leq -1$ and $n=1$, the results are summarized in Table 8. For these values of parameters $\alpha$ and $n$, the first and second NEC terms are positive for $r>0$ and $r\in (3,3.5)$. Thus, NEC and WEC are valid for   $r\in (3,3.5)$. SEC term $\rho+p_r+2p_t>0$ for all $r>0$. This shows the satisfaction of SEC for $r\in (3,3.5)$. Both DEC terms are found to be negative everywhere which means that DEC is violated everywhere. The anisotropy parameter $\triangle<0$ and the equation of state parameter $\omega>0$ for every $r>0$ which indicates the presence of attractive geometry filled with ordinary fluid. For this case, the results are summarized in Table 9\\

\noindent
\textbf{Case II(d):}  $n<0$\\
In this case, either $\alpha>0$ or $\alpha<0$. When $\alpha>0$,
$\rho>0$ for $r>1$. The first and second NEC terms are positive for $r>1$ and $r>1.6$ respectively. Thus, NEC as well as WEC are satisfied for $r>1.6$. The SEC term $\rho+p_r+2p_t$ is positive for $r\in(0,1]$. Thus, SEC is dissatisfied everywhere. Like NEC, the first and second DEC terms are positive for $r>1$ and $r>1.6$ respectively. Therefore, DEC is also valid for $r>1.6$. All NEC, WEC and DEC hold for $r>1.6$. The anisotropy parameter $\triangle<0$ for $r>1$ and $\triangle>0$ for $r\leq 1$. This shows that the geometry is repulsive near the throat and attractive away from it. The equation of state parameter $\omega>0$  for $r\in(1,1.7)$ and $-1<\omega<0$ for $r\in (0,1.04]\cup[1.67,\infty)$.  This shows the presence of non-phantom or ordinary fluid inside the wormhole. Hence, we have good results for $\alpha>0$.  At last, when $\alpha<0$, then the energy density is negative and all  energy conditions are violated. The results are also summarized in Table 10 and for $n<0$, $\alpha>0$, the results are plotted in Figures 2(a)-2(g).  \\

\noindent
Morris and Thorne\cite{morris1} constructed traversable wormhole solutions with constant redshift function in general relativity and claimed that exotic
matter is required at least at throat unless the throat may be closed. However, the study on traversable wormhole without exotic matter is a really interesting topic. Therefore, recently, several authors have tried to investigate wormhole solutions in $f(R, T)$ gravity some of them are listed here:
Zubair et al \cite{Zubair2016epjc}
investigated static spherically symmetric wormhole in $f(R, T)$ gravity by considering isotropic, anisotropic and barotropic fluids and obtained solutions are supported by non-exotic matter in few regions of the space time.
Moraes et al \cite{Moraes2017jcap}
constructed static traversable wormhole with constant throat radius in $f(R, T)$ gravity and they obtained energy conditions are satisfied for wide range of $r$. Subsequently, several authors have studied wormhole solutions in $f(R, T)$ gravity\cite{godani1, Yousaf2017epjp, Moraes2017prd, Elizalde2018prd,  Sharif2019aop, Elizalde2019prd, Zubair2019ijgmmp, Godani2019cjp}.
It is natural to compare the results in the setting of $f(R,T)$ gravity with the results in general relativity (GR). In this paper, for $\alpha=0$, the model reduces to GR. In Cases I(a) and II(a), the results are mentioned for constant and variable redshift functions respectively. In each case, all energy conditions are violated not only at throat, but also outside of the throat.  It indicates that the presence of exotic type matter is necessary to support the existence of wormhole geometries using the concept of general relativity. Furthermore, in modified gravity, the results are obtained for $n=0$, $n>0$ and $n<0$. In case of constant redshift function, the results are favorable for (a) $n=1$,  $\alpha<0$ and (b) $n<0$, an integer, $\alpha>0$.  For variable redshift function, we have desirable results for  (a) $n=1$,  $\alpha\leq -1$ and (b) $n<0$, $\alpha>0$.  Thus,  the suitable choices of $f(R)$ function, shape function $b(r)$  and redshift function have led to the favourable results confirming the existence of wormhole geometries without support of exotic matter.

\section{Conclusion}
Starobinsky \cite{Star} replaced Einstein's action $R$ with the function
$f(R) = R+\alpha R^2$ and proposed a consistent inflationary cosmological model that well explains the acceleration at early epoch of the universe and has been studied extensively in literature. In this work, we have assumed its general form $f(R)=R+\alpha R^n$, where $\alpha$ and $n$ are arbitrary constants and  studied it in the context of wormhole metric in $f(R)$ gravity. Since the wormhole metric is dependent on two arbitrary functions, namely redshift function and shape function. The redshift function can be constant or variable.
In this work, we have used the variable redshift function  $\phi(r)=\ln(\frac{r_0}{r}+1)$ \cite{Pavlovic} as well as the constant redshift function. Further, the shape function  is taken as $b(r)=\frac{r}{\exp(r-r_0)}$. The goal of the present work is to find the existence of wormhole structures containing minimum amount of  exotic matter at or near the throat and large amount of matter satisfying the energy conditions outside the throat.\\

\noindent
For each redshift function, the energy density, null, weak, strong and dominated energy condition terms, anisotropy and equation of state parameters are determined and analyzed for different possible values of parameters. For I. $\phi(r)= $ constant, NEC, WEC and DEC are observed to be validated for (i) $r>2$ with $\alpha<0$,   $n=1$; (ii) $r\in(0.4,1)\cup(1.6,\infty)$  with $\alpha>0$, $n<0$, an integer except $n=-2$; (iii) $r\in(0.3,1)$ with $\alpha>0$, $n=-2$ and (iv)$r>1.6$ with $\alpha<0$,   $n=-2$. Further, for II. $\phi(r)=\ln(\frac{r_0}{r}+1)$, NEC, WEC and SEC are obeyed for (i) $r\in(3,3.5)$ with $\alpha\leq -1$,   $n=1$ and NEC, WEC and DEC are satisfied for (ii) $r>1.6$ with $\alpha>0$, $n<0$.
For these ranges of parameters, the wormhole geometry contains exotic matter only at the throat that to a very small portion of the geometry  i.e. near the throat energy conditions are violated, matter content is phantom and geometric configuration is repulsive, however, outside the throat, the energy conditions, namely NEC, WEC and DEC, are satisfied and the matter content is non-phantom or ordinary having attractive geometric configuration.\\

\noindent
Hence, it could be concluded that the wormhole with variable redshift function is more appropriate than constant redshift function. Because, in variable redshift function case, the presence of exotic matter could be avoided by assuming the size of the throat is more than 1.6(i. e. $r>1.6$) for any $n<0$, however, in case of constant redshift function, the presence of exotic matter could be avoided by assuming the size of the throat is more than 2 (i. e. r>2) for particular choice of $n=1$. Furthermore, it is also concluded that the presence of exotic matter may not be avoided in Starobinsky model\cite{Star} with respect to the above particular choice of redshift and shape function. Therefore, redshift and shape function play as an important role for the construction of wormhole geometry not only in general relativity but also in modified gravity.
\\

\noindent
\textbf{Acknowledgement:} The authors are very much thankful to the anonymous reviewer and editor for the constructive comments for improvement of the work.

\begin{table}[!h]
	\centering
	\caption{Summary of results for $\phi(r) = constant$ with $\alpha=0$}
	\begin{tabular}{|c|c|l|}
		\hline
		S.No.& Terms& Results\\
		\hline
		1 & $\rho$ & $>0$, for $r\in(0,1)$\\
		&       & $<0$, for $r\in(1,\infty)$\\
		&       &  indeterminate, for $r=1$\\
		\cline{1-3}
		2 & $\rho+p_r$ &  $<0$, for $r\in(0,\infty)-\{1\}$\\
			\cline{1-3}
		3 & $\rho+p_t$ & $\geq0$, for $r\in (0,2]-\{1\}$\\
		&            & $<0$, $r\in(2,\infty)$\\
		&       & indeterminate, for $r=1$\\
		\cline{1-3}
			4 &  $\rho+p_r+2p_t$ & oscillating for $r\in(0, 3.8]-\{1\}$\\
					&        & indeterminate, for $r=1$ \\
		\cline{1-3}
		5 &  $\rho-|p_r|$ & $<0$, for $r\in(0, \infty)-\{1\}$\\
		&& indeterminate, for $r=1$\\
		\cline{1-3}
		6 &  $\rho-|p_t|$ & $>0$, for $r\in(0, 0.7)$ \\
		&               & $<0$, for $r\in [0.7,\infty)$-\{1\}\\
			&        &  indeterminate, for $r=1$\\
		\cline{1-3}
		7 &  $\triangle$ & $>0$, for $r\in(0, \infty)-\{1\}$ \\
		&&		indeterminate, for $r=1$\\
		\cline{1-3}
		8 &  $\omega$& $<-1$, for $r\in(0,1)$\\
		&&$>0$, for $r\in(1,\infty)$\\
		&& indeterminate, for $r=1$ \\
				\hline
		
	\end{tabular}
\end{table}

\begin{table}[!h]
	\centering
	\caption{Summary of results for $\phi(r) = constant$ with $n=0$}
	\begin{tabular}{|c|c|l|l|}
		\hline
		S.No.& Terms& $\alpha>0$& $\alpha<0$\\
		\hline
		1 & $\rho$ & $>0$, for $r\in(0,\infty)-\{1\}$&$>0$, for $r\in(0,0.5)$\\
		&       & indeterminate, for $r=1$&$<0$, for $r\in[0.5,\infty)-\{1\}$\\
		&       &  &  indeterminate, for $r=1$\\
		\cline{1-4}
		2 & $\rho+p_r$ &  $<0$, for $r\in(0,\infty)-\{1\}$& $<0$, for $r\in(0,\infty)-\{1\}$\\
		&       &  indeterminate, for $r=1$ &  indeterminate, for $r=1$\\
			\cline{1-4}
		3 & $\rho+p_t$ & $\geq0$, for $r\in (0,2]-\{1\}$&$\geq0$, for $r\in(0,2]-\{1\}$\\
		&            & $<0$, $r\in(2,\infty)$&$<0$, for  $r\in(2,\infty)$\\
		&       &  indeterminate, for $r=1$ &  indeterminate, for $r=1$\\
		\cline{1-4}
			4 &  $\rho+p_r+2p_t$ & $<0$, for $r\in(0, \infty)-\{1\}$&$>0$, for $r\in(0, \infty)-\{1\}$ \\
					&        &  indeterminate, for $r=1$ &  indeterminate, for $r=1$\\
		\cline{1-4}
		5 &  $\rho-|p_r|$ & $<0$, for $r\in(0, \infty)-\{1\}$&$<0$, for $r\in(0, \infty)-\{1\}$ \\
					&        &  indeterminate, for $r=1$ &  indeterminate, for $r=1$\\
		\cline{1-4}
		6 &  $\rho-|p_t|$ & $>0$, for $r\in(0, 2]-\{1\}$ & $<0$, for $r\in(0,\infty)-\{1\}$\\
		&               & $<0$, for $r\in (2,\infty)$&  indeterminate, for $r=1$\\
			&        &  indeterminate, for $r=1$ & \\
		\cline{1-4}
		7 &  $\triangle$ & $>0$, for $r\in(0, \infty)-\{1\}$ & $>0$, for $r\in(0, \infty)-\{1\}$\\
		&&		indeterminate, for $r=1$&	indeterminate, for $r=1$\\
		\cline{1-4}
		8 &  $\omega$& $<-1$, for $r\in(0,\infty)-\{1\}$&  $>-1$ and $<0$, for $r\in(0,\infty)-\{1\}$\\
		&& indeterminate, for $r=1$ & indeterminate, for $r=1$\\
				\hline
	\end{tabular}
\end{table}

\begin{table}[!h]
	\centering
	\caption{Summary of results for $\phi(r) = constant$ with $n=1$}
	\begin{tabular}{|c|c|l|l|}
		\hline
		S.No.& Terms& $\alpha>0$& $\alpha<0$\\
		\hline
		1 & $\rho$ & $>0$, for $r\in(0,1)$&$>0$, for $r\in(1,\infty)$\\
		&       & $<0$, for $r\in(1,\infty)$&$<0$, for $r\in(0,1)$\\
		&       & indeterminate, for $r=1$ &  indeterminate, for $r=1$\\
		\cline{1-4}
		2 & $\rho+p_r$ &  $<0$, for $r\in(0,\infty)-\{1\}$& $>0$, for $r\in(0,\infty)-\{1\}$\\
		&       &  indeterminate, for $r=1$ &  indeterminate, for $r=1$\\
			\cline{1-4}
		3 & $\rho+p_t$ & $>0$, for $r\in (0,2]-\{1\}$&$>0$, for $r\in(2,\infty)$\\
		&            & $<0$, $r\in(2,\infty)$&$<0$, for $r\in (0,2]-\{1\}$\\
		&       &  indeterminate, for $r=1$ &  indeterminate, for $r=1$\\
		\cline{1-4}
			4 &  $\rho+p_r+2p_t$ & oscillating for $r\in(0, \infty)-\{1\}$&oscillating for $r\in(0, \infty)-\{1\}$ \\
					&        &  indeterminate, for $r=1$ &  indeterminate, for $r=1$\\
		\cline{1-4}
		5 &  $\rho-|p_r|$ & $<0$, for $r\in(0, \infty)-\{1\}$& $>0$, for $r\in[2, \infty)$\\
		&& indeterminate, for $r=1$& $<0$, for $r\in(0,2)-\{1\}$\\
		&&& indeterminate, for $r=1$\\
		\cline{1-4}
		6 &  $\rho-|p_t|$ & $>0$, for $r\in(0, 0.7)$ & $>0$, for $r\in(2,\infty)$\\
		&               & $<0$, for $r\in [0.7,\infty)$-\{1\}& $<0$, for $r\in(0,2]-\{1\}$\\
			&        &  indeterminate, for $r=1$ &  indeterminate, for $r=1$\\
		\cline{1-4}
		7 &  $\triangle$ & $>0$, for $r\in(0, \infty)-\{1\}$ & $<0$, for $r\in(0, \infty)-\{1\}$\\
		&&		indeterminate, for $r=1$&	indeterminate, for $r=1$\\
		\cline{1-4}
		8 &  $\omega$& $>0$, for $r\in(1,\infty)$& $>0$, for $r\in(1,\infty)$\\
		&&  $<-1$, for $r\in(0,1)$& $<-1$, for $r\in(0,1)$\\
		&& indeterminate, for $r=1$ & indeterminate, for $r=1$\\
				\hline
		
	\end{tabular}
\end{table}

\begin{table}[!h]
	\centering
	\caption{Summary of results for $\phi(r) = constant$ with $n=2$}
	\begin{tabular}{|c|c|l|l|}
		\hline
		S.No.& Terms& $\alpha>0$& $\alpha<0$\\
		\hline
		1 & $\rho$ & $>0$, for $r\in(0,0.4)\cup(3,\infty)$&$>0$, for $r\in[0.4,2.7]$\\
		&        & $<0$, for $r\in[0.4,3]-\{1\}$&$<0$, for $r\in(0,0.4)\cup(2.7,\infty)$\\
		&        &  indeterminate, for $r=1$ &  indeterminate, for $r=1$\\
		\cline{1-4}
		2 & $\rho+p_r$ &  $>0$, for $r\in[2.3,\infty)$& $>0$, for $r\in(0,2.1)-\{1\}$\\
		&& $<0$, for $r\in(0,2.3)-\{1\}$& $<0$, for $r\in[2.1,\infty)$\\
			&        & indeterminate, for $r=1$ & indeterminate, for $r=1$\\
		\cline{1-4}
		3 & $\rho+p_t$ & $>0$, for $r\in (0,1.6]-\{1\}$&$>0$, for $r\in(1.5,\infty)$\\
		&            & $<0$, $r\in(1.6,\infty)$&$<0$, for $r\in (0,1.5]-\{1\}$\\
			&        & indeterminate, for $r=1$ & indeterminate, for $r=1$\\
		\cline{1-4}
		4 &  $\rho+p_r+2p_t$ & $>0$, for $r\in(0,3]-\{1\}$& $>0$, for $r\in(0,\infty)-\{1\}$\\
		    &&  $<0$, for $r\in(3,\infty)$&  indeterminate, for $r=1$\\
		&        &  indeterminate, for $r=1$ & \\
		\cline{1-4}
		5 &  $\rho-|p_r|$ & $>0$, for $r\in[4.6, \infty)$& $>0$, for $r\in(1,2.1]$\\
		  &            & $<0$, for $r\in(0,4.6)-\{1\}$& $<0$, for $r\in(0,1)\cup(2.1,\infty)$\\
		&& indeterminate, for $r=1$&  indeterminate, for $r=1$\\
		\cline{1-4}
		6 &  $\rho-|p_t|$ & $>0$, for $r\in(0, 0.2]$ & $>0$, for $r\in(1.5,2.6]$\\
		&               & $<0$, for $r\in [0.2,\infty)$-\{1\}& $<0$, for $r\in(0,1.5]\cup(2.6,\infty)-\{1\}$\\
		&        &  indeterminate, for $r=1$ &  indeterminate, for $r=1$\\
		\cline{1-4}
		7 &  $\triangle$ & $>0$, for $r\in(0, 2.1)-\{1\}$ & $>0$, for $r\in[2.1, \infty)$\\
		&& $<0$, for $r\in[2.1,\infty)-\{1\}$ & $<0$, for $r\in(0,2.1)-\{1\}$\\
		&&		indeterminate, for $r=1$&	indeterminate, for $r=1$\\
		\cline{1-4}
		8 &  $\omega$& $>0$, for $r\in(0.4,1.7)\cup(3.1,\infty)-\{1\}$& $>0$, for $r\in[0.4,1,6)\cup(2.9,\infty)-\{1\}$\\
		&& $ > -1$ and $<0$, for $r\in(0,0.4]\cup[1.7,2.4]$& $ > -1$ and $<0$, for $r\in(0,0.4)\cup[1.6,2.3]$\\
		&&  $<-1$, for $r\in(2.4,3.1]$& $<-1$, for $r\in(2.3,2.9]$\\
		&& indeterminate, for $r=1$ &  indeterminate, for $r=1$\\
		\hline
		
	\end{tabular}
\end{table}

\begin{table}[!h]
	\centering
	\caption{Summary of results for $\phi(r) = constant$ with  $n=6$}
	\begin{tabular}{|c|c|l|l|}
		\hline
		S.No.& Terms& $\alpha>0$& $\alpha<0$\\
		\hline
		1 & $\rho$ & $>0$, for $r\in(0,0.2)$&$>0$, for $r\in(0.1,1)$\\
		&        & $<0$, for $r\in[0.2,\infty)-\{1\}$&$<0$, for $r\in(0,0.1]\cup(1,\infty)$\\
		&        &  indeterminate, for $r=1$ &  indeterminate, for $r=1$\\
		\cline{1-4}
		2 & $\rho+p_r$ &  $<0$, for $r\in(0,\infty)-\{1\}$& $>0$, for $r\in(0,1)$\\
		&& indeterminate, for $r=1$& $<0$, for $r\in(1,\infty)$\\
		&        &  & indeterminate, for $r=1$\\
		\cline{1-4}
		3 & $\rho+p_t$ & $>0$, for $r\in (0,2)-\{1\}$&$>0$, for $r\in(0.8,2.1)-\{1\}$\\
		&            & $<0$, $r\in[2,\infty)$&$<0$, for $r\in (0,0.8]\cup[2.1,\infty)$\\
		&        & indeterminate, for $r=1$ & indeterminate, for $r=1$\\
		\cline{1-4}
		4 &  $\rho+p_r+2p_t$ & $>0$, for $r\in(0,1)\cup(1.3,2)$& $>0$, for $r\in[2,\infty)$\\
		&&  $<0$, for $r\in(1,1.3]\cup[2,\infty)$&
		$<0$, for $r\in(0,2)-\{1\}$\\
		&        &  indeterminate, for $r=1$ & indeterminate, for $r=1$ \\
		\cline{1-4}
		5 &  $\rho-|p_r|$ & $<0$, for $r\in(0, \infty)-\{1\}$& $>0$, for $r\in(0,1)$\\
		&            &indeterminate, for $r=1$  & $<0$, for $r\in(1,\infty)$\\
		&& &  indeterminate, for $r=1$\\
		\cline{1-4}
		6 &  $\rho-|p_t|$ & $>0$, for $r\in(0, 0.2)$ & $<0$, for $r\in(0,\infty)-\{1\}$\\
		&               & $<0$, for $r\in [0.2,\infty)$-\{1\}& indeterminate, for $r=1$\\
		&        &  indeterminate, for $r=1$ &  \\
		\cline{1-4}
		7 &  $\triangle$ & $>0$, for $r\in(0, \infty)-\{1\}$ & $>0$, for $r\in(0.8, \infty)-\{1\}$\\
		&& indeterminate, for $r=1$ & $<0$, for $r\in(0,0.8]$\\
		&&		&	indeterminate, for $r=1$\\
		\cline{1-4}
		8 &  $\omega$& $>0$, for $r\in(0.1,\infty)-\{1\}$& $>0$, for $r\in(0.1,\infty)-\{1\}$\\
		&&  $<-1$, for $r\in(0,0.1]$& $<-1$, for $r\in(0,0.1]$\\
		&& indeterminate, for $r=1$ &  indeterminate, for $r=1$\\
		\hline
\end{tabular}
\end{table}

\begin{table}[!h]
	\centering
	\caption{Summary of results for $\phi(r) = constant$ with $n=-0.1$}
	\begin{tabular}{|c|c|l|l|}
		\hline
		S.No.& Terms& $\alpha>0$& $\alpha<0$\\
		\hline
		1 & $\rho$ & $>0$, for $r\in(0,1)$&$>0$, for $r\in(0,0.2)$\\
		&        & $<0$, for $r\in(1,\infty)$ & $<0$, for $r\in(0.2,\infty)-\{1\}$\\
		&        &indeterminate, for $r=1$  &  indeterminate, for $r=1$\\
		\cline{1-4}
		2 & $\rho+p_r$ &  $>0$, for $r\in[0.4,1)$& $<0$ or imaginary, for $r\in(0,\infty)-\{1\}$\\
		&& $<0$, for $r\in(0,0.4)\cup(1,\infty)$&  indeterminate, for $r=1$\\
		&    &    indeterminate, for $r=1$  & \\
		\cline{1-4}
		3 & $\rho+p_t$ & $>0$, for $r\in (0,1)$ & $>0$, for $r\in(0,0.5]$\\
		&            & $<0$, for $r\in(1,\infty)$&$<0$, for $r\in (0.5,\infty]-\{1\}$\\
		&        &  indeterminate, for $r=1$& indeterminate, for $r=1$\\
		\cline{1-4}
		4 &  $\rho+p_r+2p_t$ & $<0$ or imaginary, for $r\in(0.,\infty)-\{1\}$ & $>0$, for $r\in(0,0.5]$\\
		&&  indeterminate, for $r=1$&  $<0$, for $r\in(0.5,\infty)-\{1\}$\\
		&        &   &  		indeterminate, for $r=1$\\
		\cline{1-4}
		5 &  $\rho-|p_r|$ & $<0$ or imaginary, for $r\in(0.,\infty)-\{1\}$& $<0$ or imaginary, for $r\in(0.,\infty)-\{1\}$\\
		&        &  indeterminate, for $r=1$ & indeterminate, for $r=1$   \\
		\cline{1-4}
		6 &  $\rho-|p_t|$ & $>0$, for $r\in(0,1)$ & $>0$, for $r\in(0,0.3)$\\
		&               &$<0$, for $r\in(1,\infty)$& $<0$, for $r\in[0.3,1)\cup (1,\infty)$\\
		&        &  indeterminate, for $r=1$ &indeterminate, for $r=1$  \\
		\cline{1-4}
		7 &  $\triangle$ & $>0$, for $r\in(0, 0.4)$ & $>0$, for $r\in(0, 1)$\\
		&& $<0$, for $r\in[0.4,\infty)-\{1\}$ & $<0$, for $r\in(1,\infty)$\\
		&&			indeterminate, for $r=1$&indeterminate, for $r=1$\\
		\cline{1-4}
		8 &  $\omega$& $<-1$, for $r\in(0,0.4)$& $<-1$, for $r\in(0,0.5]$\\
		&&  between -1 and 0, for $r\in[0.4,1)$& between -1 and 0, for $r\in(0.5,1)$\\
		&& $<0$ or imaginary, for $r\in(1,\infty)-\{1\}$ &$<0$ or imaginary, for $r\in(1,\infty)-\{1\}$\\
		&& indeterminate, for $r=1$ &  indeterminate, for $r=1$\\
		\hline
		
	\end{tabular}
\end{table}

\begin{table}[!h]
	\centering
	\caption{Summary of results for $\phi(r) = constant$ with  $n=-1$}
	\begin{tabular}{|c|c|l|l|}
		\hline
		S.No.& Terms& $\alpha>0$& $\alpha<0$\\
		\hline
		1 & $\rho$ & $>0$, for $r\in(0,\infty)-\{1\}$&$>0$, for $r\in(0,0.2)$\\
		&        & indeterminate, for $r=1$ & $<0$, for $r\in[0.2,\infty)-\{1\}$\\
		&        &  &  indeterminate, for $r=1$\\
		\cline{1-4}
		2 & $\rho+p_r$ &  $>0$, for $r\in(0.4,\infty)-\{1\}$& $<0$, for $r\in(0,\infty)-\{1\}$\\
		&& $<0$, for $r\in(0,0.4]$& indeterminate, for $r=1$\\
		&    &    indeterminate, for $r=1$  & \\
		\cline{1-4}
		3 & $\rho+p_t$ & $>0$, for $r\in (0,1)\cup(1.6,\infty)$&$>0$, for $r\in(0,0.4)$\\
		&            & $<0$, for $(1,1.6]$ &$<0$, for $r\in [0.4,\infty)$\\
		&        & indeterminate, for $r=1$ & indeterminate, for $r=1$\\
		\cline{1-4}
		4 &  $\rho+p_r+2p_t$ & $<0$, for $r\in(0,\infty)-\{1\}$& $>0$, for $r\in(0,\infty)-\{1\}$\\
		&&  indeterminate, for $r=1$ & indeterminate, for $r=1$\\
		\cline{1-4}
		5 &  $\rho-|p_r|$ & $>0$, for $r\in(0.3,\infty)-\{1\}$ & $<0$, for $r\in(0,\infty)-\{1\}$\\
		&               & $<0$, for $r\in (0,0.3]$& indeterminate, for $r=1$\\
		&        &  indeterminate, for $r=1$ &  \\
		\cline{1-4}
		6 &  $\rho-|p_t|$ & $>0$, for $r\in (0,1)\cup(1.6,\infty)$& $<0$, for $r\in(0,0.2]$\\
		&               &$<0$, for $(1,1.6]$& $<0$, for $r\in(0.2,\infty)$\\
		&        & indeterminate, for $r=1$  &indeterminate, for $r=1$  \\
		\cline{1-4}
		7 &  $\triangle$ & $>0$, for $r\in(0, 0.4]$ & $>0$, for $r\in(0, \infty)-\{1\}$\\
		&& $<0$, for $r\in(0.4,\infty)-\{1\}$ &indeterminate, for $r=1$ \\
		&&			indeterminate, for $r=1$&\\
		\cline{1-4}
		8 &  $\omega$& $>0$, for $r\in(1,1.7)$& $>0$, for $r\in(1,1.7)$\\
		&&  $>-1$ and $<0$, for $r\in(0,1)\cup[1.7,\infty)$& $>-1$ and $<0$, for $r\in(0,1)\cup[1.7,\infty)$\\
		&& indeterminate, for $r=1$ &  indeterminate, for $r=1$\\
		\hline
		
	\end{tabular}
\end{table}

\begin{table}[!h]
	\centering
	\caption{Summary of results for $\phi(r) = constant$ with $n=-2$}
	\begin{tabular}{|c|c|l|l|}
		\hline
		S.No.& Terms& $\alpha>0$& $\alpha<0$\\
		\hline
		1 & $\rho$ & $>0$, for $r\in(0,1)$&$>0$, for $r\in(1,\infty)$\\
		&        & $<0$, for $r\in(1,\infty)$ & $<0$, for $r\in(0,1)$\\
		&        &indeterminate, for $r=1$  &  indeterminate, for $r=1$\\
		\cline{1-4}
		2 & $\rho+p_r$ &  $>0$, for $r\in(0.3,1)$& $>0$, for $r\in(1,\infty)$\\
		&& $<0$, for $r\in(0,0.3]\cup(1,\infty)$&  $<0$, for $r\in(0,1)$\\
		&    &    indeterminate, for $r=1$  & indeterminate, for $r=1$\\
		\cline{1-4}
		3 & $\rho+p_t$ & $>0$, for $r\in (0,1.7)-\{1\}$&$>0$, for $r\in(0,0.4)\cup(1.6,\infty)$\\
		&            & $<0$, for $r\in(1.7,\infty)$&$<0$, for $r\in [0.4,1.6]-\{1\}$\\
		&        &  indeterminate, for $r=1$& indeterminate, for $r=1$\\
		\cline{1-4}
		4 &  $\rho+p_r+2p_t$ & $>0$, for $r\in(0.2,1)$& $>0$, for $r\in(0,1)$\\
		&&  $<0$, for $r\in(0,0.2]\cup(1,\infty)$&  $<0$, for $r\in(1,\infty)$\\
		&        &  indeterminate, for $r=1$ &  		indeterminate, for $r=1$\\
		\cline{1-4}
		5 &  $\rho-|p_r|$ & $>0$, for $r\in(0.3,1)$ & $>0$, for $r\in(1,\infty)$\\
		&               & $<0$, for $r\in (0,0.3]\cup(1,\infty)$&  $<0$, for $r\in (0,1)$\\
		&        &  indeterminate, for $r=1$ & indeterminate, for $r=1$   \\
		\cline{1-4}
		6 &  $\rho-|p_t|$ & $>0$, for $r\in(0,1)$ & $>0$, for $r\in(1.6,\infty)$\\
		&               &$<0$, for $r\in(1,\infty)$& $<0$, for $r\in(0,1.6]$\\
		&        &  indeterminate, for $r=1$ &indeterminate, for $r=1$  \\
		\cline{1-4}
		7 &  $\triangle$ & $>0$, for $r\in(0, 0.3]\cup(1,\infty)$ & $>0$, for $r\in(0, 1)$\\
		&& $<0$, for $r\in(0.3,1)$ & $<0$, for $r\in(1,\infty)$\\
		&&			indeterminate, for $r=1$&indeterminate, for $r=1$\\
		\cline{1-4}
		8 &  $\omega$& $>0$, for $r\in(1,1.7)$& $>0$, for $r\in(1,1.7)$\\
		&&  $>-1$ and $<0$, for $r\in(0,1)\cup[1.7,\infty)$& $>-1$ and $<0$, for $r\in(0,1)\cup[1.7,\infty)$\\
		&& indeterminate, for $r=1$ &  indeterminate, for $r=1$\\
		\hline
		
	\end{tabular}
\end{table}
\begin{table}[!h]
	\centering
	\caption{Summary of results for $\phi(r) = \log(\frac{r_0}{r}+1)$ with $n=1$}
	\begin{tabular}{|c|c|l|l|}
		\hline
		S.No.& Terms& $\alpha>-1$& $\alpha\leq -1$\\
		\hline
		1 & $\rho$ & $>0$, for $r\in(0,1]$&$>0$, for $r\in(1,\infty)$\\
		&        & $<0$, for $r\in(1,\infty)$ & $<0$, for $r\in(0,1]$\\
				\cline{1-4}
		2 & $\rho+p_r$ &  $<0$, for $r\in(0,\infty)$& $>0$, for $r\in(0,\infty)$\\
				\cline{1-4}
		3 & $\rho+p_t$ & $>0$, for $r\in (0,3]$&$>0$, for $r\in(3,3.5)$\\
		&            & $<0$, for $r\in(3,\infty)$&$<0$, for $r\in (0,3]\cup[3.5,\infty)$\\
		\cline{1-4}
		4 &  $\rho+p_r+2p_t$ & $<0$, for $r\in(0,\infty)$& $>0$, for $r\in(0,\infty)$\\
		\cline{1-4}
		5 &  $\rho-|p_r|$ & $<0$, for $r\in(0,\infty)$ & $<0$, for $r\in(0,\infty)$\\
				\cline{1-4}
		6 &  $\rho-|p_t|$ & $>0$, for $r\in(0,0.7)$ & $<0$, for $r\in(0,\infty)$\\
		&               &$<0$, for $r\in[0.7,\infty)$& \\
		\cline{1-4}
		7 &  $\triangle$ & $>0$, for $r\in(0, \infty)$ & $<0$, for $r\in(0, \infty)$\\
		\cline{1-4}
		8 &  $\omega$& $>0$, for $r\in(1,\infty)$& $>0$, for $r\in(0,\infty)$\\
		&&  $<0$, for $r\in(0,1]$ & \\
		\hline
	\end{tabular}
\end{table}

\begin{table}[!h]
	\centering
	\caption{Summary of results for $\phi(r) = \log(\frac{r_0}{r}+1)$ with $n<0$}
	\begin{tabular}{|c|c|l|l|}
		\hline
		S.No.& Terms& $\alpha>0$& $\alpha<0$\\
		\hline
		1 & $\rho$ & $>0$, for $r\in(1,\infty)$&$<0$, for $r\in(0,\infty)$\\
		&        & $<0$, for $r\in(0,1]$ & \\
				\cline{1-4}
		2 & $\rho+p_r$ &  $>0$, for $r\in(1,\infty)$& $>0$, for $r\in(1,\infty)$\\
		&        & $<0$, for $r\in(0,1]$ & $<0$, for $r\in(0,1]$\\
				\cline{1-4}
		3 & $\rho+p_t$ & $>0$, for $r\in (1.6,\infty)$&$<0$, for $r\in(0,\infty)$\\
		&            & $<0$, for $r\in(0,1.6]$&\\
		\cline{1-4}
		4 &  $\rho+p_r+2p_t$ & $>0$, for $r\in(0,1]$& $<0$, for $r\in(0,\infty)$\\
		\cline{1-4}
		5 &  $\rho-|p_r|$ & $>0$, for $r\in(1,\infty)$ & $<0$, for $r\in(0,\infty)$\\
		&               &$<0$, for $r\in(0,1]$& \\
				\cline{1-4}
		6 &  $\rho-|p_t|$ & $>0$, for $r\in(1.6,\infty)$ & $<0$, for $r\in(0,\infty)$\\
		&               &$<0$, for $r\in(0,1.6]$& \\
		\cline{1-4}
		7 &  $\triangle$ & $<0$, for $r\in(1, \infty)$ & $>0$, for $r\in(0, \infty)$\\
				\cline{1-4}
		8 &  $\omega$& $>0$, for $r\in(0,1]\cup[1.7,\infty)$& $>0$, for $r\in(0,\infty)$\\
		&&  between -1 and 0, for $r\in(1,1.7)$ & \\
		\hline
	\end{tabular}
\end{table}

\begin{figure}
	\centering
	\subfigure[Density]{\includegraphics[scale=.68]{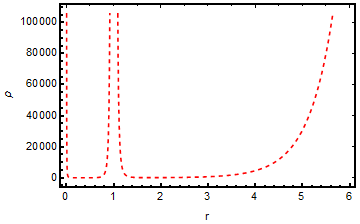}}\hspace{.05cm}
	\subfigure[NEC]{\includegraphics[scale=.68]{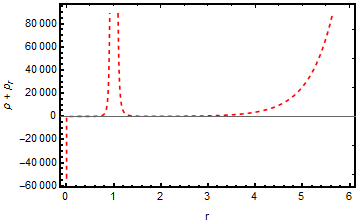}}\hspace{.05cm}
	\subfigure[NEC]{\includegraphics[scale=.68]{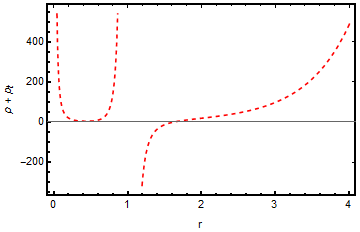}}\vspace{.05cm}
	\subfigure[SEC]{\includegraphics[scale=.68]{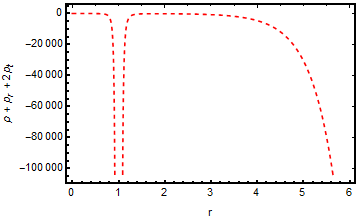}}\vspace{.05cm}
	\subfigure[DEC]{\includegraphics[scale=.68]{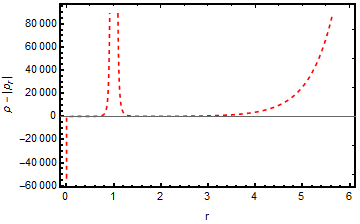}}\hspace{.05cm}
	\subfigure[DEC]{\includegraphics[scale=.68]{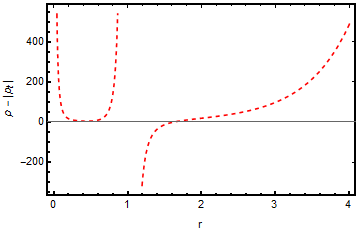}}\hspace{.05cm}
			\end{figure}
	\begin{figure}
\centering
\subfigure[$\triangle$]{\includegraphics[scale=0.68]{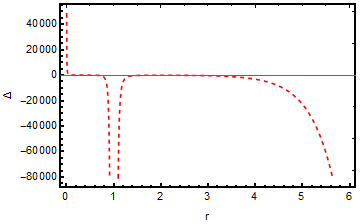}}\vspace{.05cm}
	\subfigure[$\omega$]{\includegraphics[scale=.68]{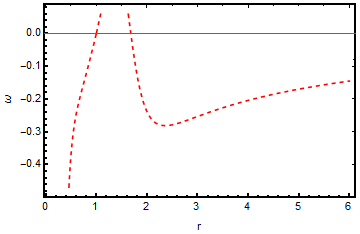}}\hspace{.05cm}
	\caption{Plots for Density, NEC, SEC, DEC, $\triangle$ \& $\omega$ using $\phi(r)$ = constant with $n=-1$ and $\alpha>0$}
\end{figure}

\begin{figure}
	\centering
	\subfigure[Density]{\includegraphics[scale=.62]{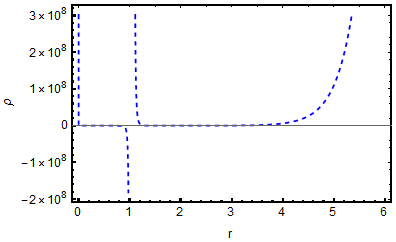}}\hspace{.05cm}
	\subfigure[NEC]{\includegraphics[scale=.62]{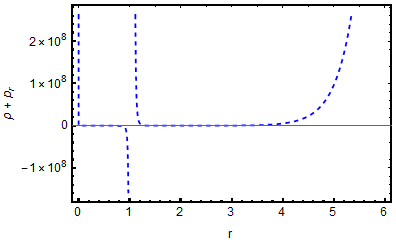}}\hspace{.05cm}
	\subfigure[NEC]{\includegraphics[scale=.62]{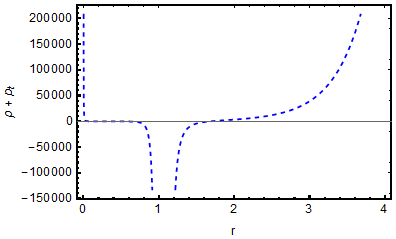}}
\vspace{.05cm}
\subfigure[SEC]{\includegraphics[scale=.62]{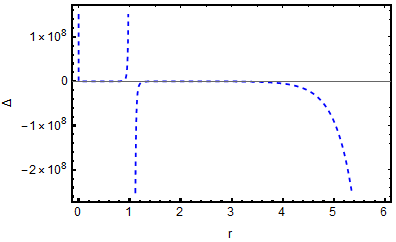}}\vspace{.05cm}
			\end{figure}
	\begin{figure}
\centering
\subfigure[DEC]{\includegraphics[scale=.62]{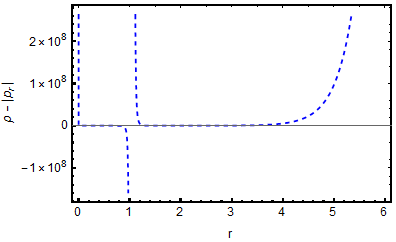}}\hspace{.05cm}
\subfigure[DEC]{\includegraphics[scale=.62]{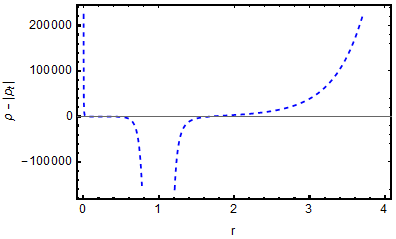}}\hspace{.05cm}
	\subfigure[$\triangle$]{\includegraphics[scale=0.62]{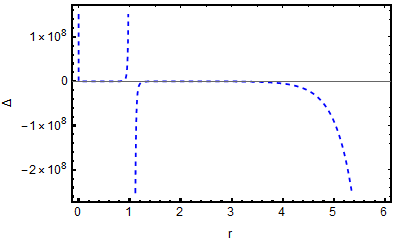}}\vspace{.05cm}
	\subfigure[$\omega$]{\includegraphics[scale=.62]{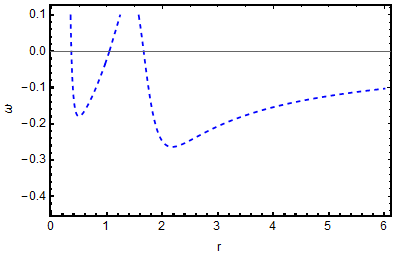}}\hspace{.05cm}
	\caption{Plots for Density, NEC, SEC, DEC, $\triangle$ \& $\omega$ using $\phi(r) = \log(\frac{r_0}{r}+1)$ with $n<0$ and $\alpha>0$}
\end{figure}

\end{document}